\documentclass[usegraphicx,usenatbib,useAMS]{mn2e}
\usepackage{times}
  \let\url\relax
\def\apj{{ApJ}}
\def\apjs{{ApJS}}
\def\apjl{{ApJL}}
\def\aap{{\em A.\&A}}

\def\mnras{{MNRAS}}

\def\prd{{Phys. Rev. D}}
\def\apss{Ap\&SS}
\def\physrep{Physics Reports}
\newcommand{\be}{\begin{equation}}
\newcommand{\ba}{\begin{eqnarray}}
\newcommand{\ee}{\end{equation}}
\newcommand{\ea}{\end{eqnarray}}  

\def\lesssim{\mathrel{\hbox{\rlap{\hbox{\lower4pt\hbox{$\sim$}}}\hbox{$<$}}}}
\def\gtrsim{\mathrel{\hbox{\rlap{\hbox{\lower4pt\hbox{$\sim$}}}\hbox{$>$}}}}
\def\gtsima{$\; \buildrel > \over \sim \;$}
\def\ltsima{$\; \buildrel < \over \sim \;$}
\def\gsim{\lower.5ex\hbox{\gtsima}}
\def\lsim{\lower.5ex\hbox{\ltsima}}
\def\simgt{\lower.5ex\hbox{\gtsima}}
\def\simlt{\lower.5ex\hbox{\ltsima}}
\def\simpr{\lower.5ex\hbox{\prosima}}

 \newcommand{\hMsun}{h^{-1}\,M_\odot}
 
\def\simless{\mathbin{\lower 3pt\hbox
   {$\rlap{\raise 5pt\hbox{$\char'074$}}\mathchar''7218$}}}   % < or of order
\def\simgreat{\mathbin{\lower 3pt\hbox
   {$\rlap{\raise 5pt\hbox{$\char'076$}}\mathchar''7218$}}}   % > or of order

\begin{document}

\title[Observability of Cosmic Reionization]{
%Reionization observables for WMAP 3-year results] {
%Reionization observables for WMAP 3-year results
Current models of the observable consequences of cosmic reionization and
their detectability
} \author[I. T. Iliev, et
al.]{Ilian~T.~Iliev$^1$\thanks{e-mail: iliev@cita.utoronto.ca}, Garrelt
  Mellema$^{2}$, Ue-Li Pen$^1$, J. Richard Bond$^1$, \newauthor
  Paul~R.~Shapiro$^3$,
  \\
  $^1$ Canadian Institute for Theoretical Astrophysics, University
  of Toronto, 60 St. George Street, Toronto, ON M5S 3H8, Canada\\
  $^2$ Stockholm Observatory, AlbaNova
  University Center, Stockholm University, SE-106 91 Stockholm, Sweden\\
  $^3$ Department of Astronomy, University of Texas, Austin, TX 78712-1083,
  U.S.A.}  \date{\today} \pubyear{2006} \volume{000} \pagerange{1} \twocolumn
\maketitle
\label{firstpage}

\begin{abstract}
A number of large current experiments aim to detect the signatures of the
Cosmic Reionization at redshifts $z>6$. Their success depends crucially on
understanding the character of the reionization process and its observable
consequences and designing the best strategies to use. We use large-scale 
simulations of cosmic reionization to evaluate the reionization signatures 
at redshifted 21-cm and small-scale CMB anisotropies in the best current 
model for the background universe, with fundamental cosmological
parameters given by WMAP 3-year results (WMAP3). We find that the optimal
frequency range for observing  the ``global step'' of the 21-cm emission  
is 120-150 MHz, while statistical studies should aim at 140-160~MHz,
observable by GMRT. Some strongly-nongaussian brightness features should 
be detectable at frequencies up to $\sim190$~MHz. In terms of
sensitivity-signal trade-off relatively low resolutions, corresponding to
beams of at least a few arcminutes, are preferable. The CMB anisotropy signal
from the kinetic Sunyaev-Zel'dovich effect from reionized patches peaks at
tens of $\mu$K at arcminute scales and has an {\it rms} of $\sim1\mu$K, and
should be observable by the Atacama Cosmology Telescope and the South Pole
Telescope. We discuss the various observational issues and the uncertainties
involved, mostly related to the poorly-known reionization parameters and, to a 
lesser extend, to the uncertainties in the background cosmology. 
\end{abstract}


\section{Introduction}\label{intro}
The Epoch of Reionization and the preceding Cosmic Dark Ages, from
recombination at $z\sim1100$ to $z\sim6$ include the formation of the first
nonlinear cosmological structures, first stars, QSO's and the emergence of the
Cosmic Web as we know it today. Yet it still remains poorly understood. This
is mostly due to the scarcity of direct observations, resulting in weak
constraints on the theoretical models. However, this situation is set to
improve markedly in the coming years, with the construction of a number of new
observational facilities, particularly for detection of the redshifted 21-cm
line of hydrogen\footnote{Giant Metrewave Radio Telescope (GMRT;
  {\url{http://www.ncra.tifr.res.in}}), Low Frequency Array (LOFAR;
  \url{http://www.lofar.org}), Murchison Widefield Array (MWA;
  {\url{http://web.haystack.mit.edu/arrays/MWA}}), Primeval Structure
  Telescope (PAST; \url{http://web.phys.cmu.edu/$\sim$past/}), and Square
  Kilometre Array (SKA; {\url{http://www.skatelescope.org}}).}, and the
kinetic Sunyaev-Zel'dovich (kSZ) effect\footnote{South Pole Telescope (SPT; 
  {$\url{http://spt.uchicago.edu/spt}$}) and Atacama Cosmology Telescope
  (ACT; $\url{http://www.physics.princeton.edu/act/}$)}.
Simulations of that epoch, required in order to make reliable predictions
for such observations, are difficult and computationally-intensive. Recently 
we presented a set of large-scale, high-resolution radiative transfer 
simulations of cosmic reionization 
\citep{2006MNRAS.369.1625I,2006MNRAS.372..679M,2007MNRAS.376..534I}. These 
simulations were the first ones which were sufficiently large to reliably 
capture the characteristic scales of the reionization process. This allowed 
us to derive the first realistic predictions for the reionization observables, 
in particular the different signatures in the redshifted 21-cm emission of 
neutral hydrogen \citep{2006MNRAS.372..679M} and the imprint of the ionized 
patches on small-scale CMB temperature anisotropies through the kinetic
Sunyaev-Zel'dovich (kSZ) effect \citep{2006NewAR..50..909I,kSZ}. We also
proposed an approach to use the obtained 21-cm maps to derive the Thomson
optical depth fluctuations due to reionization \citep{pol21} and derived the
CMB polarization signatures of patchy reionization
\citep{2007astro.ph..1784D}. We studied the effects of varying ionizing source
efficiencies and sub-grid gas clumping on these observed signals. All of these
calculations except \citet{2007astro.ph..1784D} used a particular set of
cosmological parameters, based on the best-fit first-year WMAP results, 
hereafter WMAP1, with the following cosmological parameters: 
($\Omega_m,\Omega_\Lambda,\Omega_b,h,\sigma_8,n)=(0.27,0.73,0.044,0.7,0.9,1)$
\citep{2003ApJS..148..175S}. Here $\Omega_m$, $\Omega_\Lambda$, and $\Omega_b$ 
are the total matter, vacuum, and baryonic densities in units of the critical 
density, $\sigma_8$ is the {\it rms} density fluctuations extrapolated to the 
present on the scale of $8 h^{-1}{\rm Mpc}$ according to the linear 
perturbation theory, and $n$ is the index of the primordial power spectrum of 
density fluctuations. 

Recently the 3-year WMAP results were published \citep{2006astro.ph..3449S}, 
hereafter WMAP3, which presented an updated, and fairly different best-fit 
cosmology: ($\Omega_m,\Omega_\Lambda,\Omega_b,h,\sigma_8,n)=(0.24,0.76,0.042,
0.73,0.74,0.95)$. We also note that other current measurements of these
parameters based on e.g. other CMB experiments, supernovae, large-scale
structure, clusters, and Ly-$\alpha$ forest tend to give slightly different
values, either on their own or in combination with the WMAP data 
\citep[e.g.][]{2006astro.ph..3449S,2006astro.ph..4335S,2006JPhG...33....1Y}.
In particular, they tend to yield higher value of $\sigma_8$ than WMAP3 alone 
(but still well below the WMAP1 value), at $\sigma_8\sim0.8-0.85$. For example,
recent results using all of the CMB data derives $\sigma_8=0.79$, while 
combining with the large-scale structure data yields $\sigma_8=0.81$  
\citep{acbar}. 

In this work we evaluate the detectability of reionization at radio
wavelength observations of redshifted 21-cm line of hydrogen and CMB
anisotropies from kSZ. We derive a variety of observational signatures and
discuss their detectability with a number of current and near-future
experiments and discuss the related uncertainties due to various poorly-known 
parameters. We give special attention to matching the parameters of
the observations (beamsizes, bandwidths, frequencies, observational
strategies) to the character of the reionization features in order to maximize
their detectability. We also compare our predicted signals to the expected
sensitivities for several current experiments.   

For the benefit of the reader, whenever possible we compare our current
results with our previous predictions done in the framework of the WMAP1
cosmology. The major difference between the WMAP1 and WMAP3 cosmologies is the
overall amplitude of the power spectrum, expressed here in terms of
$\sigma_8$, but the models also have slightly different spectral shapes, with
the low $\sigma_8$ one having a red tilt, $n_s-1 = -0.05$. We have previously
shown \citep{2006ApJ...644L.101A,2007MNRAS.376..534I} that these changes
result in structure formation being delayed in the WMAP3 universe relative to
the WMAP1 universe, so the epoch of reionization is shifted to lower
redshifts. In particular, if source halos of a given mass are assumed to have
released ionizing photons with the same efficiency in either case, then
reionization for WMAP3 is predicted to have occurred at $(1+z)$-values which
are roughly 1.3-1.4 times smaller than for WMAP1. The predicted
electron-scattering optical depth of the IGM accumulated since the beginning
of the EOR would have then been smaller for WMAP3 than for WMAP1 by a factor
of $(1.3-1.4)^{3/2}\sim1.5-1.7$, just as the observations of large-angle
fluctuations in the CMB polarization require. This means that the ionizing
efficiency per collapsed baryon required to make reionization early enough to
explain the value of $\tau_{\rm es}$ reported for WMAP1 and WMAP3 are nearly
the same.  

This delay of reionization can be understood in terms of the density
fluctuations at the scales relevant to reionization as follows. Let us 
denote the {\it rms} linear amplitudes on the top hat smoothing scales 
of $0.1\,h^{-1}$Mpc and $0.01\,h^{-1}$Mpc by $\sigma_{0.1}$ and
$\sigma_{0.01}$, respectively. The top hat scale $0.1\,h^{-1}$Mpc
corresponds to a mass $2.7\times 10^8 \hMsun$ for the low $\sigma_8$
case, and $3.1\times 10^8 \hMsun$ for the high $\sigma_8$ case, the
slight difference being due to the differing $\Omega_m$ and $h$. Of
course $0.01\,h^{-1}$Mpc corresponds to masses 3 orders of magnitude
smaller, $2.7\times 10^5 \hMsun$ and $3.1\times 10^5 \hMsun$, for WMAP3 
and WMAP1. Thus, $\sigma_{0.1}$ and $\sigma_{0.01}$ roughly 
correspond to the scales of the dwarf galaxies and minihaloes, 
respectively. The shape difference in the low and high $\sigma_8$ cases 
is encoded in the ratios $\sigma_{0.1}/\sigma_8$, 
6.1 and 6.6, and $\sigma_{0.01}/\sigma_8$, 10.1 and 11.0, that is, not 
negligible but not that large relative to the 20\% decrease in $\sigma_8$. 
A reasonable indication of when structure on scale $R$ formed at high
redshift is $1+z_R \approx 1.3 \sigma_R \Omega_m^{-0.23}$, where
$[a/D(z)] \approx \Omega_m^{0.23}$ for $z\gg1$. Here $D(z)$ is the 
linear growth factor from redshift $z$ to the present. 
For the minihalo scale $R=0.01\,h^{-1}$Mpc, $z_{0.01} \approx$ 12.5 and 
16.4, respectively. For the dwarf scale $z_{0.1} \approx$ 7.1 and 9.4,
respectively, in reasonable accord with the computed overlap redshifts
from our inhomogeneous reionization simulations \citep[][and 
Table~\ref{summary_wmap3} below]{2007MNRAS.376..534I}; the 50\% reionization 
redshifts bracket the minihalo and dwarf structure formation redshifts. 
The uniform reionization Thompson depth $\tau$ to a reionization redshift 
$z_{rei}$ is $\tau = 0.085[(1+z_{rei})/11]^{3/2}$ for the low $\sigma_8$ 
model and $\tau = 0.080[(1+z_{rei})/11]^{3/2}$ for the high $\sigma_8$ 
model. When the $z_{50\%}$ values are substituted, there is rough 
agreement with the $\tau_{\rm es}$ in Table~\ref{summary_wmap3}. The 
scaling of the Thompson depth $\tau$ would be $(1+z_R)^{3/2}$, about 1.4, 
roughly consistent with the $\approx 1.3$ ratio we determine, and with 
the results in \citet{2006ApJ...644L.101A} and \citet{2007MNRAS.376..534I}.

Some of the observable implications of this delay in the formation of
structures are fairly straightforward. For example, the decrease in the 
mean spatially-averaged redshifted 21-cm signal as the IGM reionizes, 
referred to as ``global step'' \citep{1999A&A...345..380S} will occur at 
lower redshift, higher frequency, in WMAP3 cosmology. It should thus become 
a bit easier to observe than previously thought, due to the lower foregrounds
and higher sensitivity at high frequencies. However, other consequences of the   
new cosmology framework are less obvious and have to be evaluated with care.

This paper is organized as follows. In \S~\ref{sims} we briefly describe 
our simulations. In \S~\ref{21cm_sect} we present our predictions for the
redshifted 21-cm signals and discuss their observability with current and
planned radio arrays. In \S~\ref{kSZ_sect} we evaluate the patchy kSZ signal 
and its observability with ACT and SPT telescopes. Our conclusions are
summarized in \S~\ref{conclusions_sect}.

\section{Simulations}\label{sims}

Our simulations were performed using a combination of two very efficient
computational tools, a cosmological particle-mesh code called PMFAST
\citep{2005NewA...10..393M} for following the structure formation, whose
outputs are then post-processed using our radiative transfer and
non-equilibrium chemistry code called C$^2$-Ray \citep{methodpaper}.  Our
simulations, parameters and methodology were discussed in 
\citet{2006MNRAS.369.1625I,2006MNRAS.372..679M} and
\citet{2007MNRAS.376..534I}. Detailed tests of our radiative transfer method
were presented in \citet{methodpaper} and \citet{comparison1}. The simulations
considered in this work are summarized in Table~\ref{summary_wmap3}, along
with the basic characteristics of their reionization histories\footnote{WMAP1
  cases listed here were first presented in \citep{2006MNRAS.372..679M}, with
  predictions of 21-cm background and kSZ effect from patchy reionization from
  those cases presented in \citet{2006MNRAS.372..679M} and \citet{kSZ},
  respectively.}. The parameter $f_\gamma$ characterizes the
emissivity of the ionizing sources - how many ionizing photons per gas atom in
the (resolved) halos are produced and manage to escape from the host halo
within $\sim20$~Myr, which is the time between two consecutive density
slices, equal to two radiative transfer timesteps. Both of our WMAP3
simulations utilize $f_\gamma=250$, as this value yields final H~II region
overlap at $z_{\rm ov}\approx6.5-7.5$ in agreement with the current observational
constraints. The corresponding integrated Thomson scattering optical depths,
$\tau_{\rm es}$ are also in agreement, within 1-$\sigma$ of the WMAP3 derived
value, $\tau_{\rm es}=0.09\pm0.03$, although they are a bit lower than the
central value. Much lower ionizing efficiencies for the sources than these
will result in too late an overlap, violating the available observational
constraints, while much higher ones would result in an early-reionization
scenario, again possibly in conflict with observations. The corresponding
WMAP1 simulations with the same ionizing source efficiencies of $f_\gamma=250$
yielded $\tau_{\rm es}=0.098-0.130$, outside of the nominal 1-$\sigma$
WMAP1-derived range $\tau_{\rm es}=0.17\pm0.04$. As we have previously shown
\citep{2007MNRAS.376..534I}, the presence of low-mass ionizing sources
[absent here, but resolved in the smaller-box simulations presented in
\citet{2007MNRAS.376..534I}] increases the total optical depth, and can easily
bring it into agreement with any value within the WMAP3 (or WMAP1,
respectively) 1-$\sigma$ range. In the same previous work 
we showed that despite this potentially dramatic effect on the integrated
Thomson optical depth, the presence of small sources has only modest effects
on the large-scale geometry of reionization, because most of these smaller
sources were strongly clustered and as a consequence become strongly
suppressed during the later stages of reionization through Jeans-mass filtering
in the ionized regions.   

\begin{table}
\caption{Simulation parameters and global reionization history results for
the simulations used in this work. 
% with WMAP3 cosmological parameters. %Box sizes are in [$\,h^{-1}$Mpc].
}
\label{summary_wmap3}
\begin{center}
\begin{tabular}{@{}llllll}\hline
%&WMAP3& \\%& WMAP1 &&\\
\hline
                          &f250   &f250C %&f250   &f250C
\\[2mm]
\hline
mesh                      &$203^3$&$203^3$ %& $203^3$   & $203^3$      
\\[2mm]
box size [$\,h^{-1}$Mpc]  &100    & 100 %&100    & 100
\\[2mm]
$f_\gamma$                &250    & 250 %& 250         & 250
\\[2mm]
$C_{\rm subgrid}$         &1      & $C(z)$ %& 1        & $C(z)$
\\[2mm]
\hline\\ 
$z_{50\%}$                &8.9    & 8.3    %& 11.7     & 11
\\[2mm] 
$z_{\rm overlap}$         &7.5    & 6.6    %& 9.3      & 8.2
\\[2mm] 
$\tau_{\rm es}$           &0.082  & 0.076  %& 0.109    & 0.098
\\[2mm]
\hline\\
\end{tabular}
\end{center}
\end{table}

\begin{figure*}
  \includegraphics[width=7.2in]{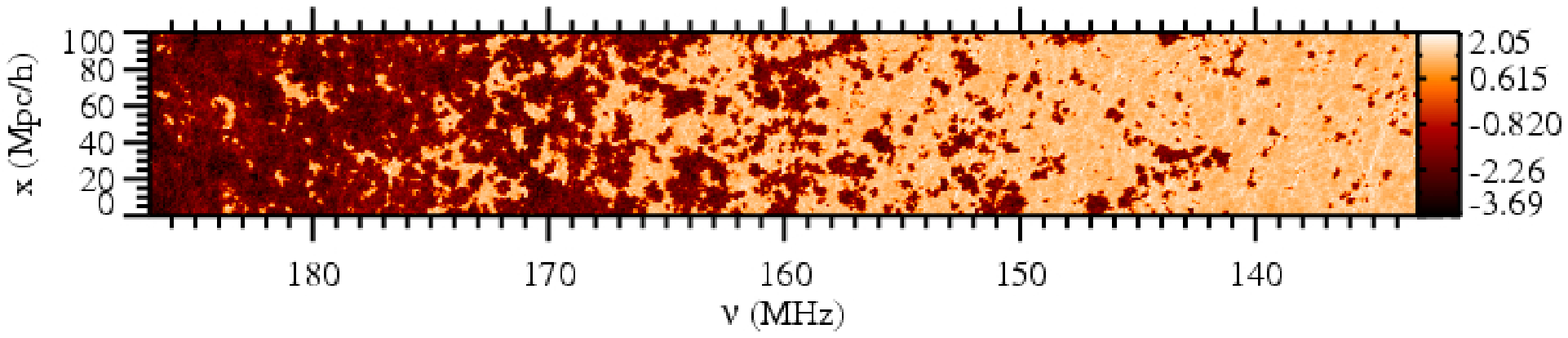}
  \includegraphics[width=7.2in]{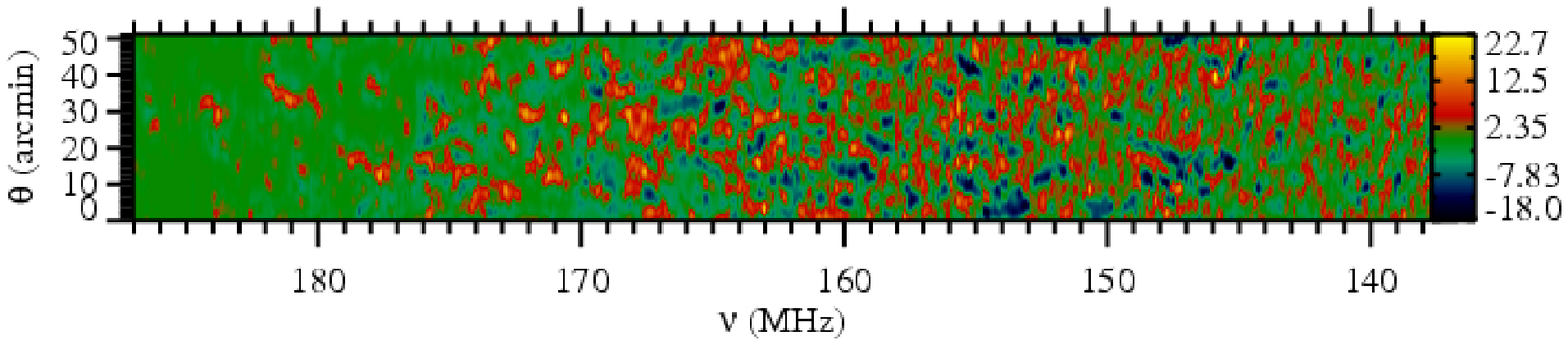}
\caption{Position-redshift slices from the f250C simulation. These slices
  illustrate the large-scale geometry of reionization and the significant 
  local variations in reionization history as seen at redshifted 21-cm line. 
  Observationally they correspond to slices through an image-frequency
  volume. The top image shows the decimal log of the differential brightness 
  temperature at the full grid resolution, while the bottom image shows the 
  same $T_b$ data, but in linear scale and smoothed with a compensated Gaussian 
  beam of 3' and (tophat) bandwidth of 0.2 MHz, roughly corresponding to the
  expected parameters for LOFAR. The spatial scale is given in comoving units
  (top) or (approximate) angle on the sky in arcminutes (bottom).
\label{pencil}}
\end{figure*}

\section{21-cm emission}
\label{21cm_sect}

The differential brightness temperature with respect to the CMB of 
the redshifted 21-cm emission is determined by the density of neutral 
hydrogen, $\rho_{\rm HI}$, and its spin temperature, $T_{\rm S}$. 
It is given by 
\ba
  \delta T_b &=&\frac{T_{\rm S} - T_{\rm CMB}}{1+z}(1-e^{-\tau})\nonumber\\
&\approx&
\frac{T_{\rm S} - T_{\rm CMB}}{1+z}\frac{3\lambda_0^3A_{10}T_*n_{HI}(z)}{32\pi
  T_S H(z)}\\
&=&{28.5\,\rm mK}\left(\frac{1+z}{10}\right)^{1/2}(1+\delta)
\left(\frac{\Omega_b}{0.042}\frac{h}{0.73}\right)\left(\frac{0.24}{\Omega_m}
   \right)^{1/2}  
\nonumber
\label{temp21cm}
\ea
\citep{1959ApJ...129..536F}, where $z$ is the redshift, $T_{\rm CMB}$ is the 
temperature of the CMB radiation at that redshift, $\tau$ is the corresponding 
21-cm optical depth, $\lambda_0=21.16$~cm is the rest-frame wavelength of the 
line, $A_{10}=2.85\times10^{-15}\,\rm s^{-1}$ is the Einstein A-coefficient,
$T_*=0.068$~K corresponds to the energy difference between the two levels, 
$1+\delta={n_{\rm HI}}/{ \langle n_H \rangle}$ is the mean number density 
of neutral hydrogen in units of the mean number density of hydrogen at
redshift $z$, 
\ba
\langle n_H \rangle(z)&=&\frac{\Omega_b\rho_{\rm
    crit,0}}{\mu_Hm_p}(1+z)^3\nonumber\\ 
&=&1.909\times10^{-7}\rm cm^{-3}\left(\frac{\Omega_b}{0.042}\right)(1+z)^3,
\ea
with $\mu_H=1.32$ the corresponding mean molecular weight (assuming 32\% He
abundance by mass), and $H(z)$ is the redshift-dependent Hubble constant,
\ba
  H(z)&=&
H_0[\Omega_{\rm m}(1+z)^3+\Omega_{\rm k}(1+z)^2+\Omega_\Lambda]^{1/2}
                      \nonumber\\  
&=&H_0E(z)\approx H_0\Omega_{\rm m}^{1/2}(1+z)^{3/2},
\ea
where $H_0$ is its value at present, and the last approximation is valid for 
$z\gg 1$. Throughout this work we assume that all of the neutral IGM gas is 
Ly-$\alpha$-pumped and sufficiently hot (due to e.g. a small amount of X-ray 
heating) above the CMB temperature and is thus seen in emission. These 
assumptions are generally well-justified, except possibly at the earliest
times \citep[see e.g.][and references therein]{2006PhR...433..181F}. 

\subsection{The progress of reionization: global view} 
The images in Figure~\ref{pencil} show the progress of reionization as
seen at 21-cm emission vs. the observed frequency for run f250. The technique
we used to produce them was described in detail in \citet{2006MNRAS.372..679M}
and is similar to the standard method for making light-cone images. In
short, it involves continuous interpolation between the single-redshift
numerical outputs in frequency/redshift space. The slices are cut at an
oblique angle so as to avoid repetition of the same structures along the same
line-of-sight. Redshift-space distortions due to peculiar bulk velocities are
included. The top image shows the decimal logarithm of the differential
brightness temperature at the full resolution of our simulation data,
approximately 0.25 arcmin in angle and 30~kHz in frequency. The ionization
starts from the highest density peaks, which is where the very first galaxies
form. These high peaks are strongly clustered at high redshifts, which results
in a quick local percolation of the ionized regions. As a result, by $z\sim9$
($\nu_{\rm obs}\sim140$~MHz) already a few fairly large ionized regions form,
each of size $\sim10$~Mpc. These continue to grow and overlap until there is
only one topologically-connected H~II region in our computational volume,
which for this particular simulation occurs at $z\sim7$ ($\nu_{\rm
  obs}\sim180$~MHz). However, even at this time quite large, tens of Mpc
across, neutral regions still remain. They are gradually ionized as time goes
on, but some of them persist until the very end of our simulation. 
The uncertaities in the reionization parameters (e.g. source efficiencies, gas
clumping) result in moderate variations in the typical sizes of the ionized
and neutral regions, but do not change the basic characteristics of the
reionization process. 

\begin{figure}
  \includegraphics[width=3.2in]{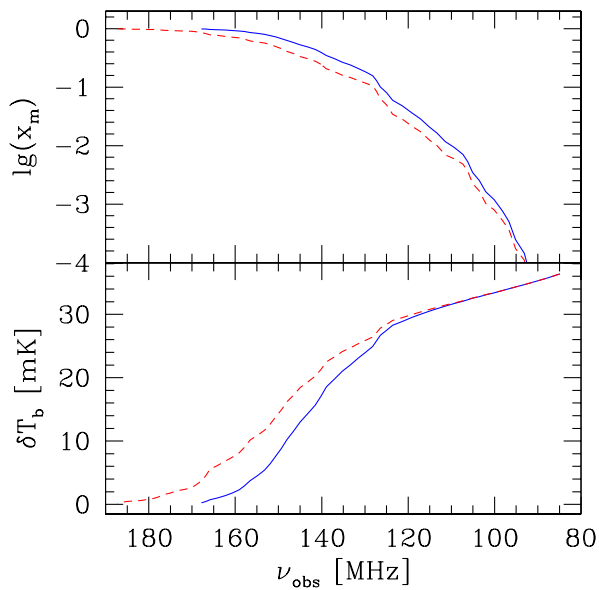}
\caption{Evolution of the mean mass-weighted ionized fraction, $x_m$, (top), 
  mean flux in $\mu$~Jy (assuming a circular beam with an angular diameter
  of 10') (middle) and the mean differential brightness temperature, $\delta
  T_b$, in mK vs. observed frequency for simulations f250 (solid, blue), and
  f250C (short-dashed, red).
\label{meantbzfig}}
\end{figure}

The bottom image in Figure~\ref{pencil} shows the same data as the top, but 
as would be seen by a radio interferometer array assuming perfect foreground 
removal. To obtain it we smoothed the data with a compensated Gaussian beam 
with FWHM of 3' and bandwidth of 0.2 MHz, similar to the expected parameters 
for the LOFAR array. Unless otherwise stated, throughout this paper we use a
compensated Gaussian beam. We recognise that a compensated Gaussian is not
perfect match to the actual interferometer beam, however it captures its
essential properties. In particular, it has zero mean, and negative troughs at
the side of the central peak \citep[see e.g.][]{2006MNRAS.372..679M}. As a
result, ionized regions show as negative differential brightness temperature 
regions if they are surrounded by neutral volumes. If an ionized (or a
neutral) region is much larger than the smoothing scale, the resulting signal  
would be close to zero. A direct comparison between the two images shows that
all the main structures clearly remain also in the smoothed image, indicating
that LOFAR and other similar interferometers would have sufficient resolution
to determine the large-scale reionization morphology to a reasonable accuracy
throughout most of the reionization history. However, at the earliest stages
of reionization some of the existing H~II regions are barely seen, or not at
all, since the beam smoothing either merges them with other nearby ionized
regions, or simply smoothes them away. This is a consequence of the small
sizes of these early H~II regions, which puts the majority of them below the
beam resolution. At frequencies higher than $\sim140$~MHz the ionized bubbles
become large enough to be above the smoothing scale and thus all major
structures become visible. Higher ionizing source efficiencies and/or lower 
gas clumping would yield somewhat larger H~II regions, thus they would become
visible slightly earlier. We also note that the Cosmic Dark Ages and the early
stages of reionization might be still observable through other sources of
21-cm fluctuations which we do not consider here. These include e.g. the 21-cm
emission and absorption by cosmological minihaloes
\citep{2002ApJ...572L.123I,2002ApJ...579....1F,2003MNRAS.341...81I}, by
shock-heated IGM \citep{2004ApJ...611..642F,2006ApJ...646..681S}, or due to
inhomogeneous early backgrounds in Ly-$\alpha$
\citep{2005ApJ...626....1B,2006ApJ...648L...1C} or X-rays
\citep{2006astro.ph..7234P}.  
We note that there are bright and potentially observable features even for
$\nu_{\rm obs}>180$~MHz. This is important for the planned observations  
since at such high frequencies both the foregrounds and the Radio Frequency 
Interference (RFI) are substantially lower, while at the same time the array 
sensitivities improve. We discuss these issues in more detail in
\S~\ref{obs_21cm}. 

In Figure~\ref{meantbzfig} we show the evolution of the mean (i.e. averaged
over a large volume) mass-weighted ionization fraction, $x_m$, and the 
mean differential brightness temperature as seen at the observer, 
$\overline{\delta T_b}(\nu_{\rm obs})$. 
As the universe steadily becomes ever more ionized, the mean 21-cm
differential brightness temperature naturally decreases. However, the signal
persists at a non-trivial level (a few mK or more) until quite late,up to
frequencies of 150-170 MHz. Similarly to our previous results which  
used the WMAP1 parameters, the ``global step'' from mostly-neutral to
mostly-ionized medium \citep{1999A&A...345..380S} turns out to be rather
gradual, with the signal decreasing by $\sim20$~mK over 20-30~MHz (and
somewhat more gradual for case f250C than for f250). The differential
brightness temperature scales with the reionization redshift as $\delta
T_b\propto(1+z)^{1/2}$. Thus, a delay of reionization from WMAP1 to WMAP3
model by 1.3 in redshift corresponds to an expected $\delta T_b$ decrease by a
factor of 1.14, roughly as found in our simulations. 

Analytical models of the globally-averaged 21-cm signal
\citep[e.g.]{2005MNRAS.363..818S,2006MNRAS.371..867F} predict evolutions which
are in rough agreement with ones we find. The models considered in
\citet{2005MNRAS.363..818S} yielded a somewhat faster evolution of the global
signal, while the wider range of models presented in
\citet{2006MNRAS.371..867F} included cases of both faster and of more gradual
evolution, depending on the details of the evolution of the Ly-$\alpha$ and
X-ray backgrounds.  

\subsection{The statistical measures}
\label{stat_sect}

\begin{figure}
  \includegraphics[width=3.5in]{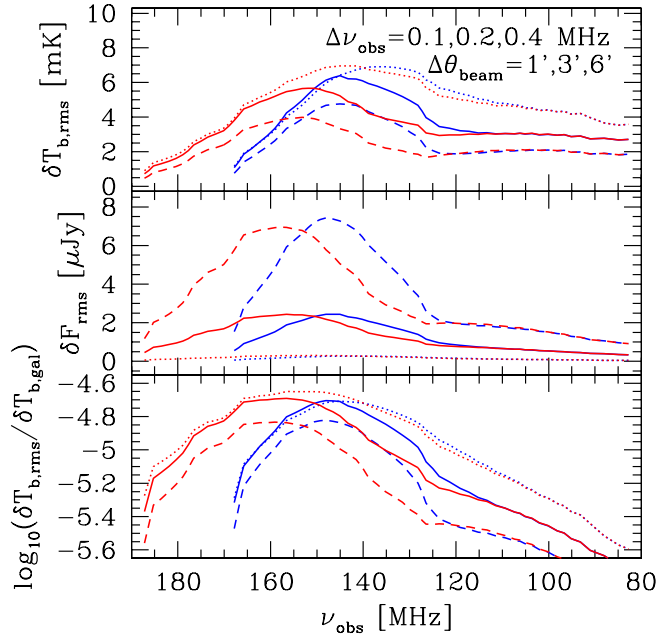}
\caption{(a)(top) {\it rms} fluctuations of the differential brightness 
  temperature, $\delta T_{\rm b,rms}\equiv\langle \delta T_b^2\rangle^{1/2}$
  vs. observed frequency, $\nu_{\rm obs}$ for f250 (blue) and f250C (red) for
  beam sizes and bandwidths $\Delta\theta_{\rm beam},\Delta\nu_{\rm
    bw})=$(1',0.1 MHz) (dotted), (3',0.2 MHz) (solid) and (6',0.4 MHz)
  (dashed); (b)(middle) Fluxes in $\mu Jy$ corresponding to the differential
  brightness temperature fluctuations in (a), same notation as in (a); and
  (c)(bottom) Ratio of $\delta T_{\rm b,rms}$ to the mean Galactic synchrotron
  foreground, assumed to be $\delta T_{\rm b,gal}=300\,\rm
  K(\nu/150\,MHz)^{-2.6}$, same notation as in (a).
\label{fluctfig}}
\end{figure}

An alternative approach to take would be a statistical one, through detection
of the fluctuations of the emission around its mean value 
\citep[see][and references therein]{1990MNRAS.247..510S,1997ApJ...475..429M,
2002ApJ...572L.123I,2004ApJ...608..622Z,2004ApJ...615....7M,
2006MNRAS.372..679M,2006PhR...433..181F}. These fluctuations
are due to a combination of the reionization patchiness and the variations of
the underlying density field. In Figure~\ref{fluctfig} we show the {\it rms}
of the 21-cm emission fluctuations derived from our simulations. The top panel
shows the evolution of the differential brightness temperature {\it rms},
$\delta T_{\rm b,rms}$, for three choices for the beam size (in arcmin; using
compensated Gaussian beam) and the bandwidth (in MHz), from top to bottom,
$(\Delta\theta_{\rm beam},\Delta\nu_{\rm bw})=(0.1,1)$ (roughly as expected
for the SKA compact core), $(0.2,3)$ (LOFAR) and $(0.4,6)$ (GMRT, MWA).  
The middle panel shows the evolution of the corresponding fluxes, given by 
\be
\delta F(\nu_{\rm obs})=\frac{2\nu_{\rm rec}^2}{c^2}
k_B\delta T_b(\nu_{\rm obs}) \Delta\Omega, 
\ee 
where $\Delta\Omega=\pi(\theta/2)^2$ is the 3D angle subtended by the beam for
a circular beam with FWHM of $\theta_{\rm beam}$. On the 
bottom panel we show the 21-cm signal as a fraction of the dominant foreground,
the Galactic synchrotron emission. For the larger beams/bandwidths the 
$\delta T_{b,rms}$ peak is at $z\sim8.8$ ($\nu_{\rm obs}\sim145$~MHz) for f250 
and at $z\sim8.3$ ($\nu_{\rm obs}\sim152-153$~MHz to for f250C, close to the 
time at which 50\% of the mass is ionized, as was the case also for our WMAP1 
results \citep{2006MNRAS.372..679M}. However, for the higher resolution,
corresponding to a $1'$ beam, the peak of the fluctuations moves to noticeably
earlier times/lower frequencies, to $\sim137$~MHz ($x_m=0.26$) for f250 and to
$\sim143$~MHz ($x_m=0.26$) for f250C. This high-resolution case differs from
the rest because the scales probed by such a small beam/bandwidth combination
are generally below the characteristic bubble size. The exception is at early
times, when ionized bubbles are still small on average, and thus match better
the smaller beam-size, which is reflected in the fluctuation peak moving to
earlier times.
\begin{figure}
  \includegraphics[width=3.3in]{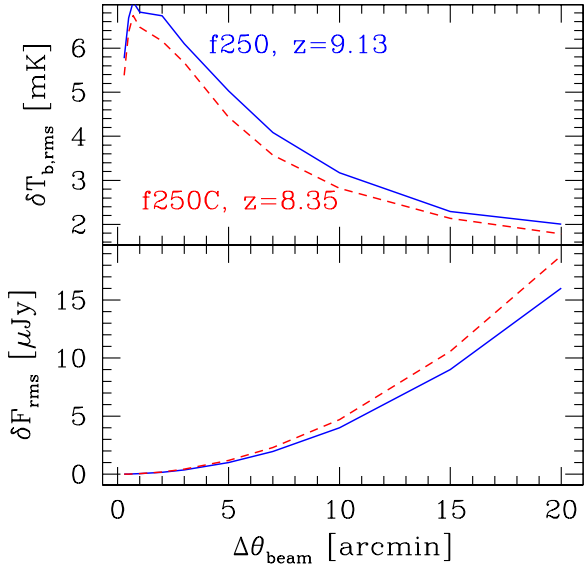}
\vspace{-0.6cm}
\caption{(a)(top) {\it rms} fluctuations of the differential brightness 
   temperature, $\delta T_{\rm b,rms}\equiv\langle \delta T_b^2\rangle^{1/2}$
   vs. beam-size, $\delta\theta_{\rm beam}$ for f250 (solid, blue) and f250C
   (short-dashed, red) at redshifts close to the maximum of the fluctuations, 
    as indicated. The bandwidth is changed in proportion to the beamsize.
    (b)(bottom) Flux fluctuations corresponding to the differential
    brightness temperature fluctuations in (a), same notation as in (a).
\label{fluctfig_beam}}
\end{figure}

\begin{figure*}
  \includegraphics[width=3.2in]{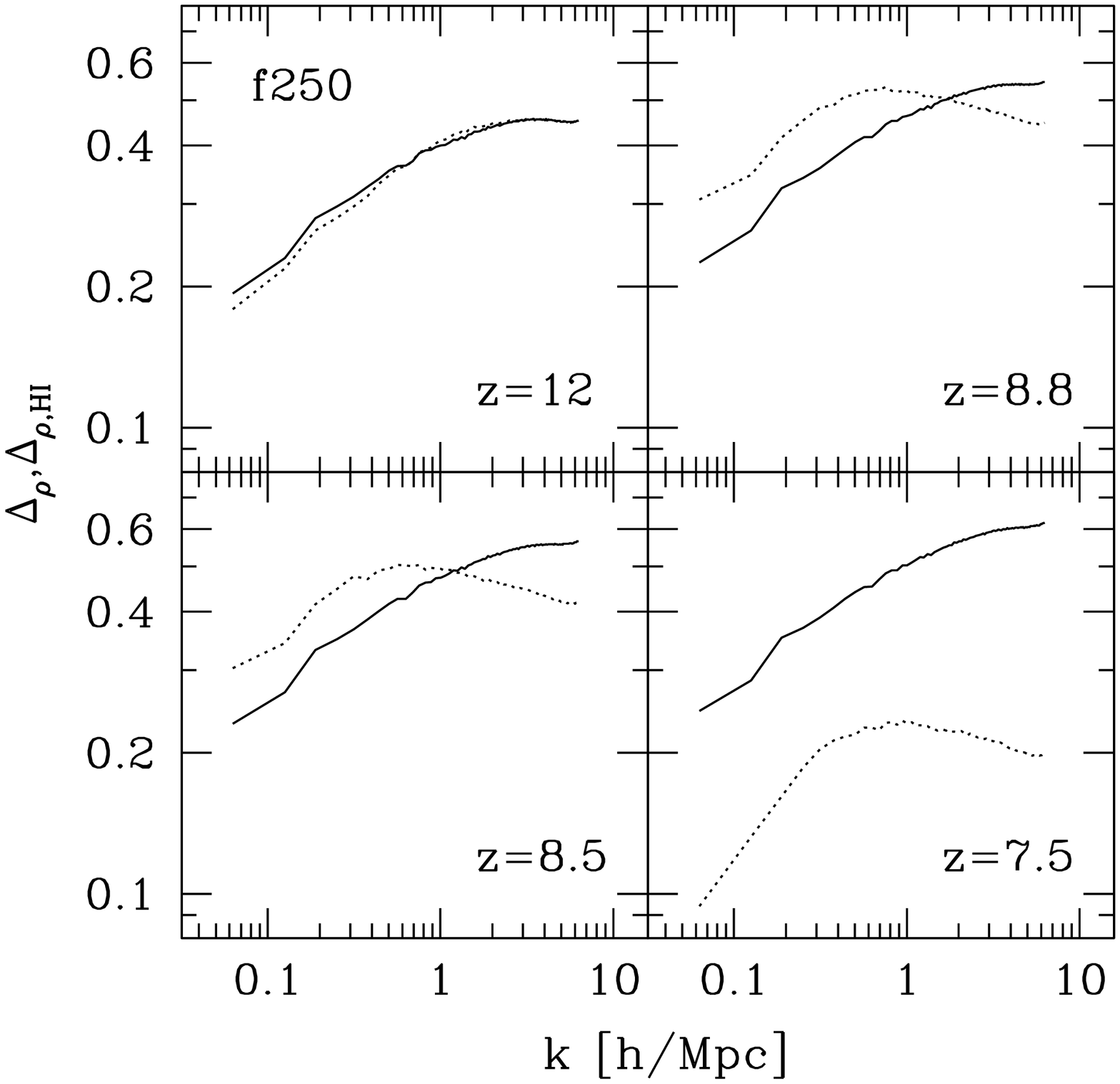}
  \includegraphics[width=3.2in]{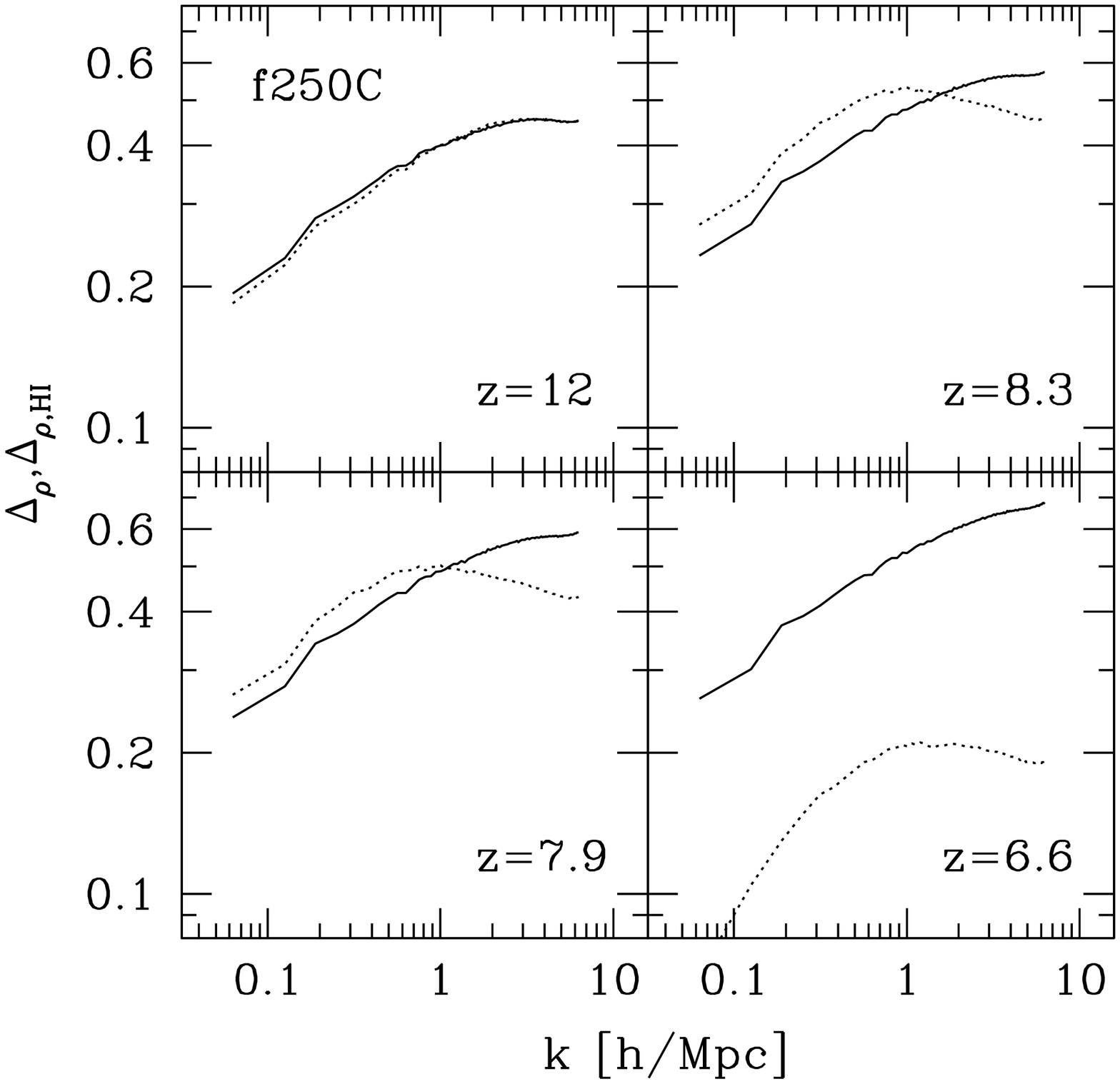}
\vspace{-0.6cm}
\caption{The variance, $\Delta$ of the 3D power spectra of the neutral 
  gas density $\Delta_{\rho,HI}$ (dotted), and the total density $\Delta_{\rho}$ 
  (solid), normalized to the total, for simulations f250 (left) and f250C
  (right). The redshifts are chosen as follows (in decreasing numerical
  order): early times ($z=12$; $\overline{T_b}=31.8$ mK for f250,
  $\overline{T_b}=33.2$ mK for f250C), the redshift at which $\delta T_{\rm
    b,rms}$ peaks ($z=8.8$, $\overline{T_b}=13.1$ mK for f250, $z=8.3$,
  $\overline{T_b}=11.8$ mK for f250C), the redshift at which $\delta F_{\rm
    rms}$ peaks ($z=8.5$, $\overline{T_b}=8.9$ mK for f250; $z=7.9$,
  $\overline{T_b}=8.2$ mK for f250C), and the redshift of overlap  
  ($\overline{T_b}=0.3$~mK for f250; $\overline{T_b}=0.2$ mK for f250C).
\label{pow3d_fig}}
\end{figure*}

We note that the flux fluctuations peak somewhat later than the corresponding 
temperature fluctuations. The position of the peak flux in
redshift/frequency space is largely independent of the resolution employed,
and is at $\sim148-150$~MHz for f250 and at $\sim155-157$~MHz for f250C. As
the beam-size and bandwidth increase from 1' to 3' and then to 6', the
temperature fluctuations decrease, albeit only by a modest amount. For
example, the peak fluctuations for $(\Delta\theta_{\rm beam},\Delta\nu_{\rm
  bw})=$(6',0.4 MHz) are only a factor of $\sim2$ lower (4 mK vs. 8 mK), while
the corresponding flux increases by over an order of magnitude, indicating
that it would be optimal to observe at relatively large scales, where we
maximize the sensitivity without sacrificing much of the signal. 

Higher source efficiency and lower gas clumping would result in the bubbles
at the same redshift being a bit larger in typical size. This would move the
peak position to a bit lower frequency. It would also make it slightly higher
for the larger beam sizes since larger bubble sizes at the half-ionized point
would match these beamsizes better. 

The 21-cm temperature fluctuations as fraction of the dominant foreground, the 
Galactic synchrotron emission (bottom panel) peak even later than the flux
fluctuations. This is due to the broad peak of the 21-cm emission and the
steep decline of this foreground at higher frequencies, which combine to push
the peak to later times/higher frequencies, at $\nu_{\rm obs}$ up to
160-165~MHz. The signal is strongly dominated by the foregrounds at all times,
but up to an order of magnitude could be gained for observations aimed close
to the peak ratio, as opposed to earlier or later times. Furthermore,
observing with an interferometer removes significant part of the foregrounds,
due to the differential nature of the measurements. This eliminates the
uniform component of the foregrounds, leaving only its fluctuations, at the
level of 1-10\% of the total.   

In Figure~\ref{fluctfig_beam} we plot the differential brightness temperature
fluctuations and the corresponding fluxes vs. the instrument smoothing, as
given by the beam size, with the bandwidth changed in proportion to the beam
size. At small scales (below the typical ionized bubble size, a few arcmin or
less) the temperature fluctuations are fairly large ($>6$ mK) and dominated by
Poisson fluctuations (e.g. a cell is either ionized or else is
neutral). However, the corresponding flux is tiny, below 1 $\mu Jy$. Around
the typical bubble scale ($\sim5-10'$) the fluctuations are slightly lower, at
$\sim3-4$~mK, but the flux grows strongly, as $\delta\theta_{\rm beam}^2$ and
reaches $\sim5\mu Jy$ at $\delta\theta_{\rm beam}=10'$. At even larger
angles/bandwidths the ionization fluctuations gradually start contributing
ever less to the total and in the large-scale limit the temperature
fluctuations just follow the underlying large-scale density fluctuations
(multiplied by the mean differential brightness temperature). The flux curve
should gradually flatten and at some point there would be little to be gained
by a further increase of the observed sky patch since the gain in angle is
canceled by the decrease of the temperature fluctuations. Thus, again there is
a trade-off between the signal level and the array sensitivity. The optimal
scale for observations will depend on the best sensitivity which could be
achieved by a particular radio array. For compact arrays the optimal beam size
is around 10-20', but could be lower than that for sensitive arrays with large
collecting area like SKA. 

Finally, in Figure~\ref{pow3d_fig} we show the three-dimensional power spectra
of the density and the neutral gas density fluctuations (to which the 21-cm 
signal is directly proportional) at several illustrative redshifts. Initially, 
the neutral density power largely tracks the one of the density field, since 
most of the gas is still neutral. At intermediate and late times the ionization
fraction inhomogeneities introduce a peak around the characteristic scale of
the ionized patches (which is $k\sim1\,h\,Mpc^{-1}$ for f250C, and slightly
lower for f250). Around the time when the fluctuations reach their peak the
patchiness boosts the power on large scales by factor of $\sim2$ (1.5) for
f250 (f250C) compared to the density power spectrum. The signal largely
disappears around the time of overlap, since little neutral hydrogen remains,
but the power spectra still show the characteristic peak, at approximately the
same scales.  

Some of the most visible redshifted 21-cm features would be the
points of maximum departure of the signal from the mean. The magnitude of the
signal is dependent on the level of smoothing and could evolve with redshift. 
The maxima/minima are roughly independent of redshift, within factor of $\sim
2$, but there are also some interesting and nontrivial features. For the case
of no smoothing (i.e. at full grid resolution, $\sim0.25'$, 30 kHz) the
maximum amplitude is quite high,  
at $\sim100-200$~mK. Naturally, the beam smoothing decreases the amplitude, to
$\sim20-30$~mK for $(\Delta\theta_{\rm beam},\Delta\nu_{bw})=(1',0.1\,\rm
MHz)$, 10-20~mK for $(\Delta\theta_{\rm beam},\Delta\nu_{bw})=(3',0.2\,\rm
MHz)$ and to 8-20~mK for $(\Delta\theta_{\rm beam},\Delta\nu_{bw})=(6',0.4\,\rm
MHz)$. The introduction of sub-grid clumping (f250 vs. f250C) results in only
minor variations here.  The absolute value of the minimum is similar to the
one for the maximum in all cases, but the two still show some differences. For 
the larger beams/bandwidths the maxima peak around the time of 50\% ionization
by mass, while the absolute values of the minima peak noticeably earlier. This
is readily understood based on the evolution of the typical ionized and neutral
region sizes and the properties of the compensated Gaussian beam. Early-on the
ionized regions are small and isolated, surrounded by large neutral patches,
which yields deep negative minima at the positions of the ionized bubbles. At
late times the H~II regions grow large and the beam-smoothed signal inside
them is close to zero. The positive maxima, on the other hand, are due to the
densest neutral regions and are sampled by the central maximum of the beam,
and thus reach their peak later.
\begin{figure}
\begin{center}
  \includegraphics[width=3.5in]{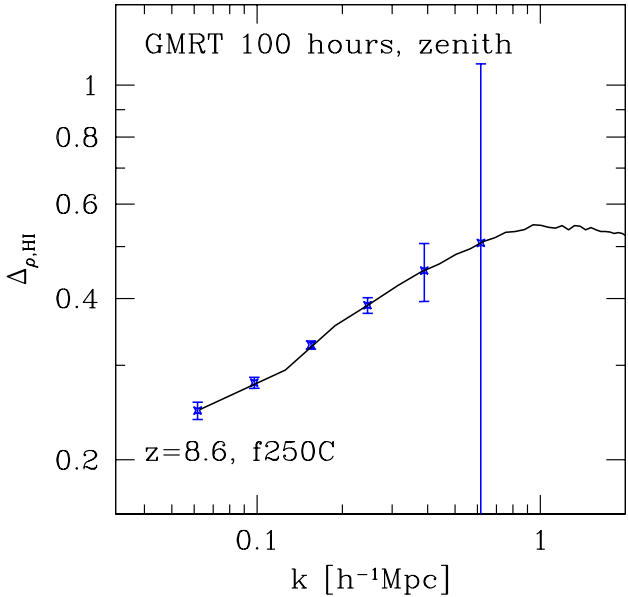}
\vspace{-0.6cm}
\caption{Observability of the 21-cm signal: the 3D power spectrum of the
  neutral hydrogen density, $\Delta_{\rho,HI}$, at redshift $z=8.59$ 
($\overline{T_{b}}=16.3$~mK) with the forecast error bars for 100 hours
observation with GMRT vs. wavenumber $k$. We assumed 15 MHz observing
bandwidth (the full instantaneous bandwidth of GMRT), $T_{\rm sys}=480$~K and
$T_S \gg T_{\rm CMB}$. The array configuration is assumed pointed to
the zenith, but the sensitivity is only weakly dependent on the pointing. 
\label{gmrt}}
\end{center}
\end{figure}

The statistics of these emission peaks is also of considerable interest since 
it shows how common such features are and thus what patch of sky one needs to
study for a detection. 
PDFs of the 21-cm emission are similar to the ones we previously found for the 
WMAP1 cosmology \citep{2006MNRAS.369.1625I}. They are considerably non-Gaussian, 
especially at late times and smaller smoothing scales. In particular, for
smoothing scales of $\sim5-10$~Mpc there is an over-abundance of the brightest
regions by up to an order of magnitude compared to the a Gaussian with the
same mean and {\it rms}. At large scales ($\simgt20\,h^{-1}$~Mpc) and late
times the PDFs become very close to Gaussian.

\subsection{Observability: redshifted 21-cm}
\label{obs_21cm}

\begin{figure}
\begin{center}
  \includegraphics[width=3.4in]{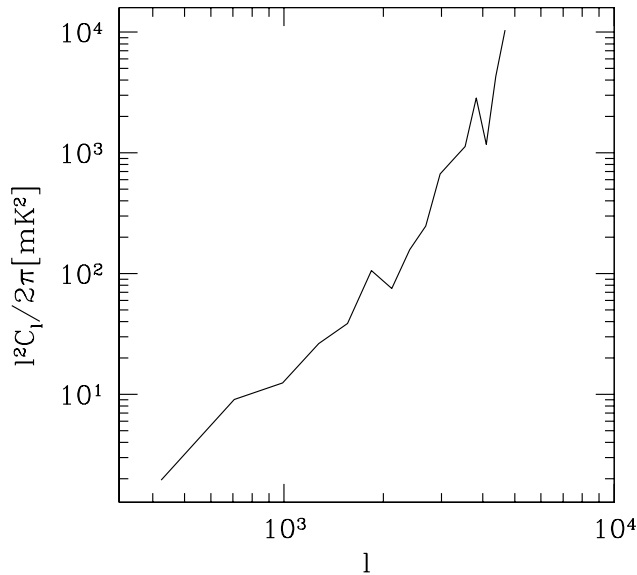}
\vspace{-0.6cm}
\caption{Observability of the 21-cm signal: forecast angular power spectrum of
  thermal noise of GMRT for bandwidth of 15 MHz, 100 hours of integration, one
  single pointing. 
\label{beamnoise_gmrt}}
\end{center}
\end{figure}

There are several current or upcoming experiments which aim to detect the
redshifted 21-cm signatures of reionization, including LOFAR, MWA, GMRT,
PAST/21CMA and SKA. Not all details of the design and the instruments are yet
known, particularly for SKA for which even the basic concept is not yet
finalized. The only arrays currently in operation are GMRT and
PAST/21CMA. Among these interferometers, GMRT and LOFAR have the largest
collecting area, at $\sim50,000\,\rm m^2$ for GMRT and (effective)
$\sim10^5\,\rm m^2$ for LOFAR at 110-200 MHz, and thus in principle they have
the best sensitivity before the commissioning of SKA. However, the same two
arrays also have significant interference problems to overcome from
terrestrial sources of confusion. As an example, in Figure~\ref{gmrt} we show
our predicted 21-cm signal at $z=8.6$ (case f250C) along with the anticipated
GMRT sensitivity to the 3D power spectrum for 100 hours of integration. The
sensitivity is calculated using the baseline distribution as seen from zenith,
angle averaged, and optimally weighted. Radially, the velocity resolution has
been assumed higher than any scale of interest, which is achievable with the
recently-developed new software correlator. The error on modes are anisotropic
for radial versus azimuth and the errors are combined by weighting each mode
by its sum of noise and signal variance. We assumed 15~MHz observing bandwidth
(the full instantaneous bandwidth of the GMRT correlator, the frequency
resolution is much better, of order kHz) and $T_{\rm sys}=480$~K. The errors
calculation assumes Gaussian noise and Gaussian source statistics, and a 7
square degree field of view. The two
small-$k$ bins would probably not be observable due to foreground removal. We
note that previous literature concentrated on two dimensional noise models for
experiments, and for reference in Figure~\ref{beamnoise_gmrt} we show the
expected band-averaged azimuthal GMRT thermal noise power spectrum for 
bandwidth of 15 MHz, 100 hours of integration, and a single pointing. Noise
is estimated using the standard radio telescope noise estimation formulae:
\be
T_{\rm map}=\frac{T_{\rm sys}}{B\sqrt{\delta \nu \delta t}},
\ee
where $B$ is the angular-scale dependent effective beam response obtained by 
summing over all the baselines assuming the telescope is pointed at the
zenith.  It changes only slightly when looking elsewhere. The beamnoise then
is simply the Fourier transform of $T_{\rm map}$. The other two arrays,
PAST/21CMA ($\sim10^4\,\rm m^2$ effective area) and MWA ($\sim7000\,\rm m^2$)
are significantly smaller than either LOFAR or GMRT, but are also more compact
and would be constructed in areas with very low interference. Thus, they could
also be quite competitive, at least in terms of detecting 21-cm fluctuations
at very large scales, because of their low spatial resolution.  

Our results indicate that the 21-cm fluctuations in WMAP3 cosmology peak
around  $\nu_{\rm obs}=130-170$~MHz (compared to 90-120 MHz for WMAP1 models),
depending on the resolution and the detailed reionization parameters (ionizing
source efficiencies and gas clumping). The corresponding 21-cm flux
fluctuations, and the 21-cm temperature fluctuations as fraction of the
foregrounds peak at even 
higher frequencies. These properties significantly facilitate the signal
detection as compared to the WMAP1 cases, since at higher frequencies the
detector sensitivities are dramatically better, while at the same time the
foregrounds are lower, by roughly factor of $\sim(120/170)^{2.6}=2.5$. The
peak value of the differential brightness temperature fluctuations is
$\sim4-8$~mK, decreasing modestly with increasing beam and bandwidth. The
corresponding fluxes are much more strongly dependent on the scale of
observations, ranging from $<1\mu\,\rm Jy$ for 1' beam and 0.1 MHz bandwidth
to $\sim8\mu\,\rm Jy$ for 6' beam and 0.4 MHz bandwidth. This clearly argues
for relatively large beams, of at least a few arcmin, and for  
maximum collecting area concentrated in the compact core, in order to improve 
the sensitivity while not losing a significant fraction of the signal. For
example, the expected sensitivity of LOFAR for 1 hour of integration time is 
$\sim0.6 \,\rm mJy$ for bandwidth of 1~MHz (and somewhat worse for the virtual
core only), while its resolution is 3.5''-6'' for the whole array and 3'-5'
for the core (numbers are frequency-dependent). Thus, only the virtual core
would be useful for EOR observations (although the extended part would be
important for e.g. better foreground subtraction). Upwards of $100$ hours of
integration time would be needed for detection, less if more antennae are
placed in the virtual core.

\begin{figure*}%[!ht]
\includegraphics[width=3.3in]{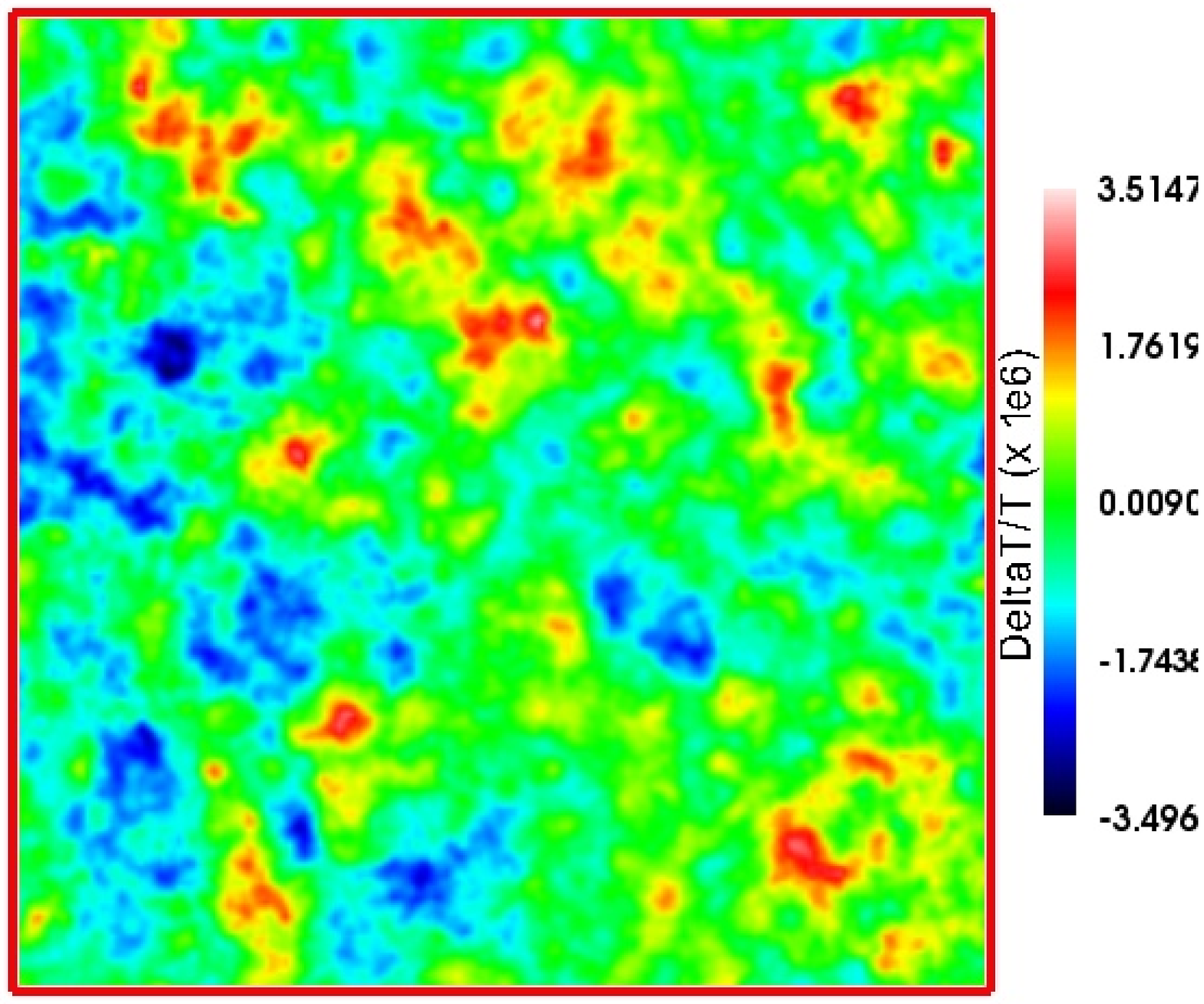}
\includegraphics[width=3.3in]{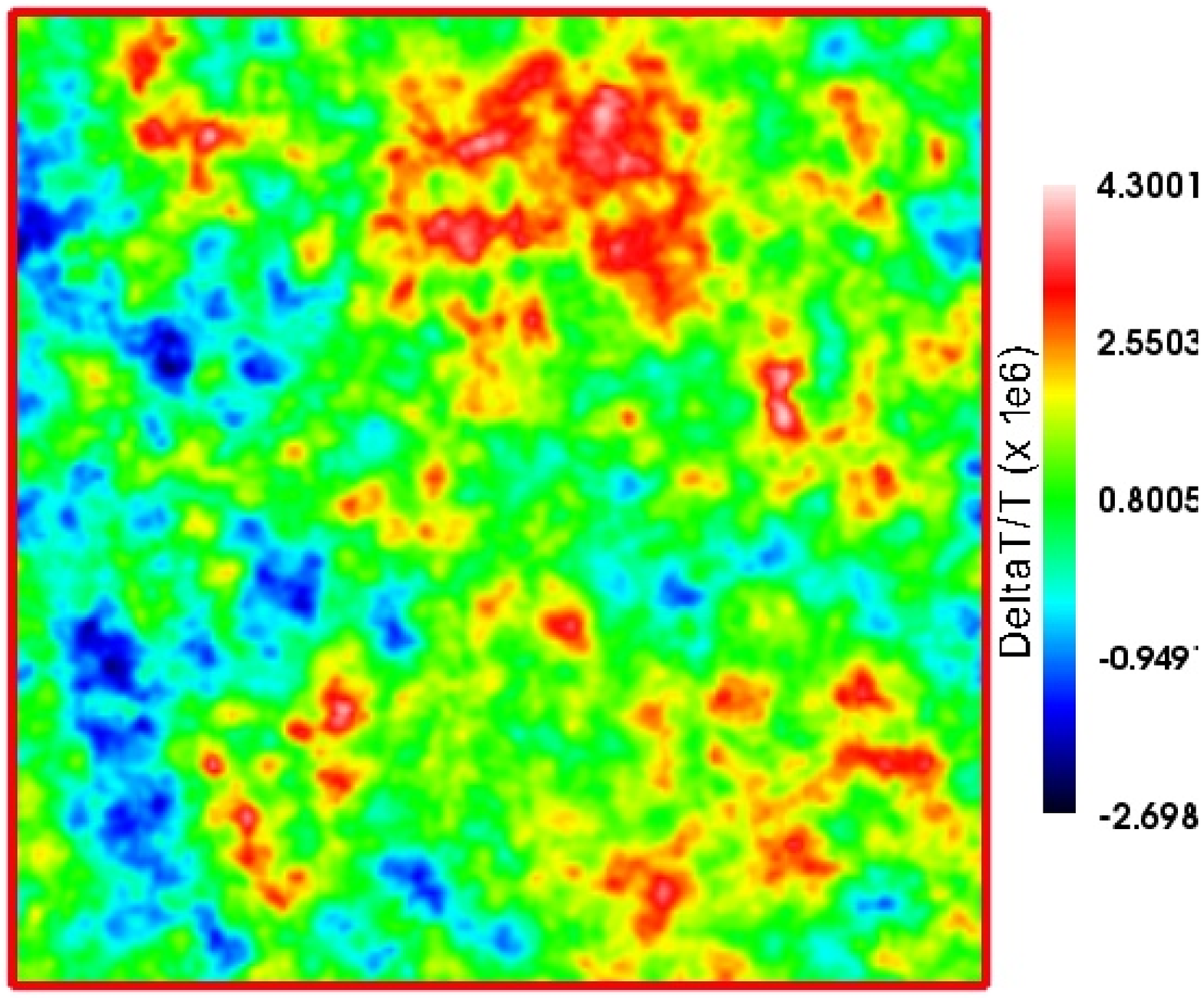}
\includegraphics[width=3.3in]{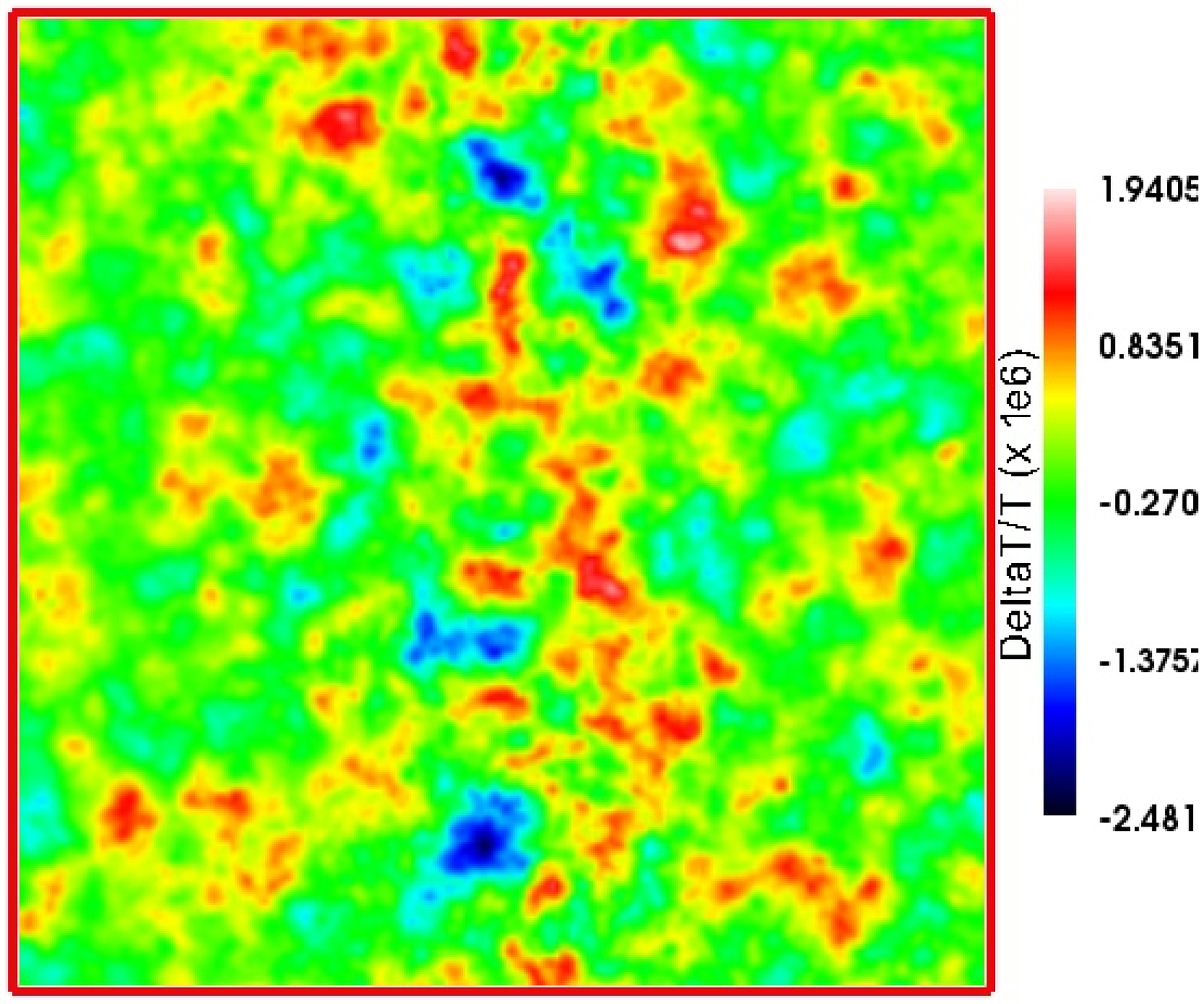}
\includegraphics[width=3.3in]{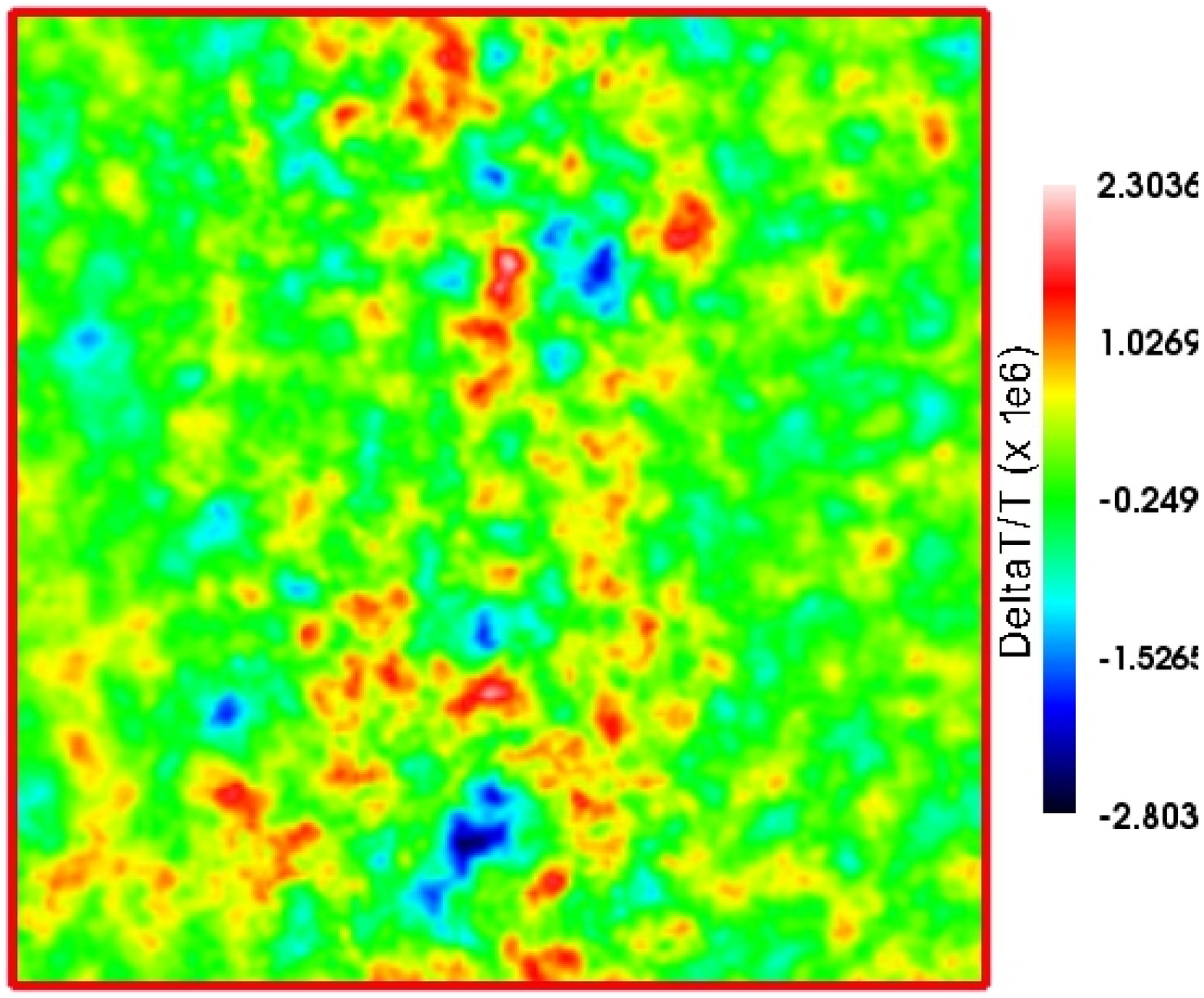}
\caption{\label{maps} kSZ temperature fluctuation maps from simulations: f250 
with large-scale velocities (top left), f250C with large-scale velocities (top 
right), f250 (bottom left), and f250C (bottom right). (Images produced using 
the Ifrit visualization package of N. Gnedin). 
}
\end{figure*}

The importance of aiming at the appropriate frequencies and length scales
cannot be overstated. As we have shown, around the peak of the fluctuations
the signal as fraction of the dominant foreground is up to an order of
magnitude larger than it is off the peak. Utilizing an appropriate beam,
i.e. one which is well-matched to the scale of the expected fluctuations and
corresponding to sufficiently large flux level, given the available
sensitivity, is also extremely important. Our results indicate that the best
beam sizes are of at least a few arcminutes, and possibly much larger, of
order 5-10' or more. Larger ionizing source efficiencies or lower gas
clumping would change the height and position of the peak fluctuations
slightly, but our main conclusions in terms of the best observation strategies
and frequency band to observe would remain valid. Once more observation data
becomes available the reionization scenarios presented here can be refined to
give us better constraints on the ionizing sources. 

Another observational strategy, and one of the first to be proposed 
\citep{1999A&A...345..380S} is to aim to detect the ``global step'' over the 
whole sky which reflects the transition from the fully-neutral universe before 
reionization to the fully-ionized one after \citep[see][for more detailed
discussion of the various observational issues]{1999A&A...345..380S}. Since
this is a global, all-sky signal it imposes essentially no requirements in
terms of resolution. The sensitivity requirements are also relatively modest,
since the flux corresponding to such a large area on the sky is very
large. The main difficulty is the foreground subtraction. The global step we
find is relatively gradual, $\sim20$~mK over $\sim20$~MHz. This would still be
readily detectable in absence of foregrounds. How well the foregrounds could
be subtracted would depend strongly on their properties, and in particular how
fast and by how much the local slope of the power law describing it
changes. If the foregrounds are well-fit by a single power law over the
relevant frequencies, then there is a good chance to detect the global
transition signal, see \citet{1999A&A...345..380S} for more detailed
discussion. 

The brightness temperature behaviour we find suggests
that the best frequency range to aim for could be $\sim120-150$~MHz, since
there the signal decrement is largest. A slightly higher overall normalization
of the primordial density fluctuation power spectrum, as suggested by
combining all the available data sets would shift the best frequency range to
slightly lower frequencies, but ones which are still above the FM range
($\nu>110$~MHz). Only a fairly high normalization, $\sigma_8\sim0.9$ and no
power spectrum tilt (as given by WMAP1) would shift the global step into the
FM range ($90\,\rm MHz<\nu<110$~MHz). However, such a high normalization is
largely excluded by the best current fits to the cosmological parameters.

We also note that albeit at first sight it appears that modifying the
background cosmology is degenerate with assuming different source
efficiencies, in fact the two options are not equivalent. A different
background cosmology changes the timing of reionization to earlier or later,
i.e. is simply a shift in redshift due to the delay or acceleration of
structure formation, as previously pointed out in
\citet{2006ApJ...644L.101A}. On the other hand, changing the source
efficiencies has a somewhat different effect - either extending or shortening  
the {\it duration} of reionization - for given power spectrum parameters the
sources form at the same time and same numbers, the different efficiencies
just mean that their effect is stronger or weaker, thereby extending or
shortening reionization, as we have discussed in \citet{2006MNRAS.372..679M}. 
While this is a somewhat subtle distinction, it does yield different 
shapes of $dT_b(z)$, so the two cases can in principle be distinguished 
and are thus not equivalent.

An important reionization signature to look for is the one due to individual, 
rare, bright features. The magnitude of such features depends on the scale 
observed, ranging from $\sim0.1$~K for high resolution (0.25', 30 kHz), to few 
tens of mK for beams of size a few arcminutes. The peak magnitude is only
weakly dependent on redshift, with the peak value within a factor of two of
the typical values. The redshift/frequency at which the peak is reached is 
beamsize-dependent, moving to higher frequencies for larger beams. Taking 
also into account the much larger fluxes corresponding to larger beams, this
again clearly argues for utilizing relatively large beams (5'-10' or more) and 
aiming at the high frequencies ($\sim130-160$~MHz) for such observations. The 
statistics of such rare, bright peaks are also favourable at late times, we
find that at $\sim3'-5'$ scales there are up to an order of magnitude more
such peaks than a Gaussian statistics would predict.

\section{kSZ effect from patchy reionization}
\label{kSZ_sect}

Next, we turn our attention to the second main reionization observable, 
the small-scale secondary CMB anisotropies generated through the kinetic 
Sunyaev-Zel'dovich (kSZ) effect \citep{1969Ap&SS...4..301Z,1980MNRAS.190..413S,
1986ApJ...306L..51O,1987ApJ...322..597V,1998PhRvD..58d3001J,1998ApJ...508..435G,
2000ApJ...529...12H,2001ApJ...551....3G,2001ApJ...549..681S,2002PhRvL..88u1301M,
2003ApJ...598..756S,2004MNRAS.347.1224Z,2005ApJ...630..657Z,2005ApJ...630..643M,
2005MNRAS.360.1063S,kSZ}. The kSZ effect results from Thomson scattering of  
the CMB photons onto electrons moving with a bulk velocity $v$. Along a
line-of-sight (LOS) defined by a unit vector ${\bf n}$ the kSZ temperature
anisotropies are given by: 
\be 
\frac{\Delta T}{T_{\rm CMB}} =\int
d\eta e^{-\tau_{\rm es}(\eta)}an_e\sigma_T \frac{{\bf n} \cdot {\bf v}}{c},
\label{ksz_int}
\ee
where $\eta=\int_0^t dt'/a(t')$ is the conformal time, $a$ is the scale
factor, $\sigma_T=6.65\times10^{-25}\,\rm cm^{-2}$ is the Thomson scattering
cross-section, and $\tau_{\rm es}$ is the corresponding optical depth. Since 
our simulation volume is not large enough to follow the complete photon
light path from high redshift to the end of reionization, we have to combine 
several simulations volumes. In order to avoid artificial effects from the box
repetition (in particular having the same structures repeating over and over 
along the photon path) we randomly shift the simulation box along both
directions perpendicular to the ray direction and also alternate the x-, y-
and z-directions of the box. We have presented our detailed methodology along
with our predictions for the kSZ reionization signal based on the WMAP1
cosmology in \citet{kSZ}. Some of the power from the largest-scale velocity
field perturbations is missing from the simulation data, due to our finite box
size. We compensate for that missing power by adding it in a statistical way,
again as discussed in detail in \citet{kSZ}. 

\subsection{The kSZ signal}

The resulting kSZ maps are shown in Figure~\ref{maps}. The full scale of the
maps corresponds to our full simulation box size ($100\,h^{-1}$~Mpc, or 
$\sim50'$) and the pixel resolution corresponds to the simulation cell size
($100\,h^{-1}/203\,\rm Mpc\approx0.5\,h^{-1} Mpc$, or $\sim0.25'$). At a few
arcminute scale there are a number of fluctuations larger than 5~$\mu$K with
both positive and negative sign. The f250 and f250C runs yield largely similar
level of fluctuations, slightly larger ones in the latter case. The typical
scale of the kSZ temperature fluctuations in run f250 is also a bit larger
than the scale for f250C, reflecting the larger, on average, sizes of the  
ionized regions in the former case compared to the latter. Introducing the 
correction for the missing large-scale velocity power increases the
fluctuation range by about 50\% and introduces some large-scale coherent
motions.      

\begin{figure}
  \includegraphics[width=3.5in]{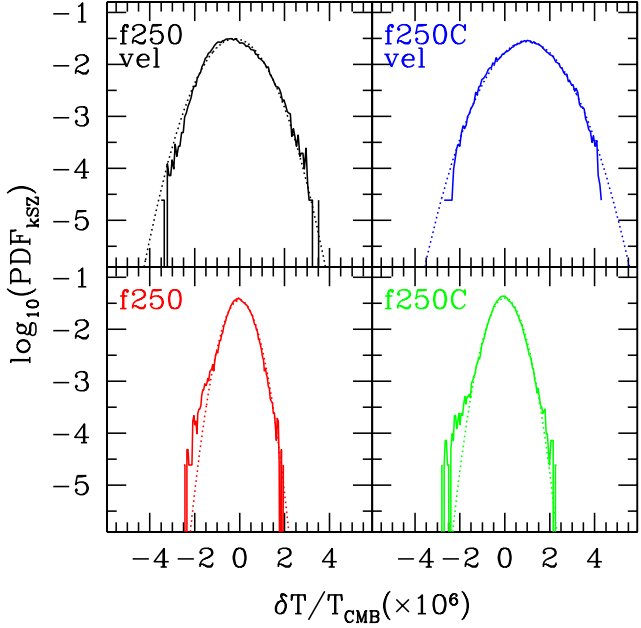}
\caption{\label{pdf} PDF distribution of $\delta T_{\rm kSZ}/T_{\rm CMB}$ 
(solid) vs. Gaussian distribution with the same mean and width (dotted) with 
(top panels) and without (bottom panels) correction for the large-scale 
velocity power missing from the computational box for our simulations, as 
labelled.}
\end{figure}

\begin{figure}
  \includegraphics[width=3.5in]{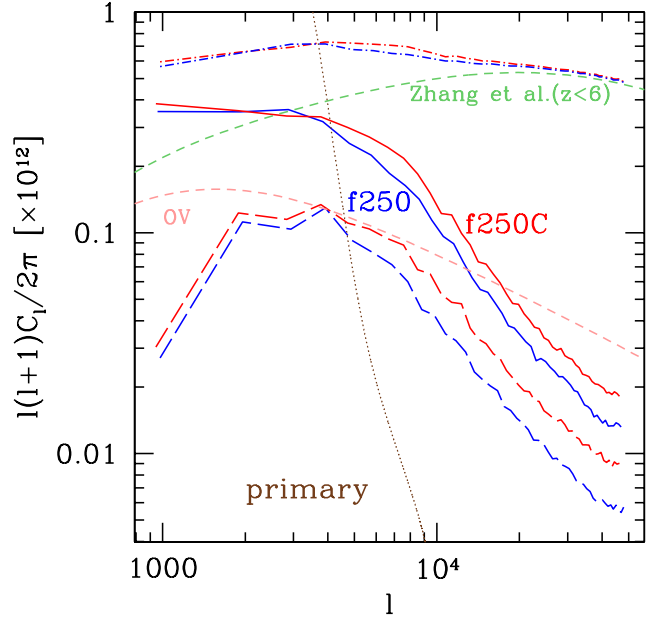}
\vspace{-0.6cm}
\caption{\label{ps} Sky power spectra of $\delta T_{\rm kSZ}/T_{\rm CMB}$
  fluctuations resulting from our simulations: f250 (blue), and f250C (red).
  Solid (dashed) lines show the results with (without) the correction for the
  large-scale velocity power missing from the box. compared to the
  after-reionization kSZ signals (assuming overlap at $z_{\rm ov}=8$): linear
  Ostriker-Vishniac effect, labelled 'OV' (long-dashed, pink), and a
  fully-nonlinear model matched to high-resolution hydrodynamic simulations of
  \citet{2004MNRAS.347.1224Z}, labelled `Zhang et al' (short-dashed, dark
  green). We also show the total (patchy reionization + Zhang et al. 
  post-reionization) signals (dot-dashed; top lines) and the primary CMB 
  anisotropy signal (dotted, brown).}
\end{figure}

In Figure~\ref{pdf} we show the PDF of these kSZ maps. The full range of 
the temperature fluctuations at pixel level is $\pm20\,\mu$K
($\pm13\,\mu$K) with (without) the large-scale velocity corrections. 
The {\it rms} fluctuations of $\Delta T/T_{\rm CMB}$ for run f250 are 
$4.835\times10^{-7}$ ($8.968\times10^{-7}$) without (with) large-scale 
velocities, while for run f250C the numbers are $5.094\times10^{-7}$ 
($1.020\times10^{-6}$) without (with) large-scale velocities. Both the 
range and the {\it rms} of the kSZ fluctuations are lower than the
corresponding quantities we derived from the WMAP1 reionization scenarios by
factor of $\sim2$, thus the PDF distributions are correspondingly less
wide. In both WMAP3 cases there are some mild departures from Gaussianity at
the bright end. Adding the correction for the large-scale velocities yields
wider (by factor of $\sim2$) and, interestingly, much more Gaussian PDF
distributions.  

Since the kSZ effect is a product of electron density and velocity, a
naive expectation is that the kSZ angular power spectrum would scale
as $\sigma_8^2$ for the density and $\sigma_v^2$ for the velocity,
$\sigma_8^4$ in total. The actual ratio we find is $\sigma_8^{3.94}$
for reionization scenario f250C and $\sigma_8^{3.83}$ for f250.
However since $8\,h^{-1}$Mpc is well above the scales relevant to
reionization, thus in the presence of tilt a more appropriate scaling
would be a bandpower at dwarf galaxy scales, $\sigma_{0.1}^2$,
introduced in \S~\ref{sims}, rather than at galaxy cluster scales,
$\sigma_{8}^2$. We find the kSZ power scales as one power less,
$\propto \sigma_{0.1}^{2.74}$ for f250C and $\propto
\sigma_{0.1}^{2.82}$ for f250. This scaling also holds true for the
minihalo bandpower, $\sigma_{0.01}^2$. Thus it becomes insensitive to
the exact length scale once we are in the relevant part of the power
spectrum. As we mentioned in \S~\ref{intro}, the values of
$\sigma_{0.01}$ and $\sigma_{0.1}$ scaled to include the growth
inhibition between high redshift and redshift zero, $(a/D) \approx
\Omega_m^{0.23}$, 0.74 and 0.76 for the WMAP3 and WMAP1 cases, provide
a good indication when the hierarchy spanning the collapse of first
the minihaloes then the dwarfs developed. 

The output from radio telescopes is sky power spectra, shown in
Figure~\ref{ps}. The kSZ signal from patchy reionization dominates the primary
anisotropy at small scales, for $\ell>4000$. The magnitude of the signal is
$[\ell(\ell+1)C_\ell/2\pi]\sim10^{-13}$, or $\Delta T\sim 1\,\mu$K. The
presence of sub-grid gas clumping boosts the power by $\sim50\%$ for
$\ell>4000$, but has little effect on the power spectrum shape.  Adding the
correction for the missing velocity power at large-scales boosts the signal
power by additional factor of 2-3 in the interesting range of scales
($\ell>4000$). At larger scales the boost is larger, by up to an order of
magnitude, but at those scales the kSZ temperature anisotropy is strongly
dominated by the primary CMB.  Compared to the predicted post-reionization
anisotropy signals, the kSZ from patchy reionization is larger than the linear
Ostriker-Vishniac signal (OV) for $\ell<20,000$ and slightly lower  
than, but similar to the full nonlinear post-reionization effect prediction by 
\citet{2004MNRAS.347.1224Z} (which we rescaled to the current WMAP3 cosmology 
using the scaling $C_\ell\propto\sigma_8^5$). The total, patchy reionization 
and post-reionization (based on \citet{2004MNRAS.347.1224Z}) signals are also
shown at the top of the figure. The total power spectrum retains the peak at
$\ell\sim3000-4000$ imprinted by the patchy reionization component, although
the signal decrease at small scales is not as pronounced, since the decrease is 
partially compensated for by the continuing weak rise of the post-reionization 
component of the signal. The two reionization scenarios (red and blue top 
curves) would be very difficult to distinguish based solely on the total
signal, but this might be possible to do if the post-reionization component is
known sufficiently well and is subtracted to a good precision (see also the
next section). A slightly higher power spectrum normalization, as suggested by
recent cosmological parameter estimates, will moderately increase the signal
as indicated by the above scalings. However, the strongly-peaked shape of the
signal will persist, since it is reflecting the characteristic scales of the
reionization process. Higher source efficiencies would shift the peak some
towards larger scales, reflecting the larger typical size of the bubbles (due
to the combination of the higher source efficiencies and the stronger source
clustering at higher redshift) in this case.   

\subsection{Observability: kSZ from patchy reionization}
\begin{figure*}
\begin{center}
  \includegraphics[width=3.3in]{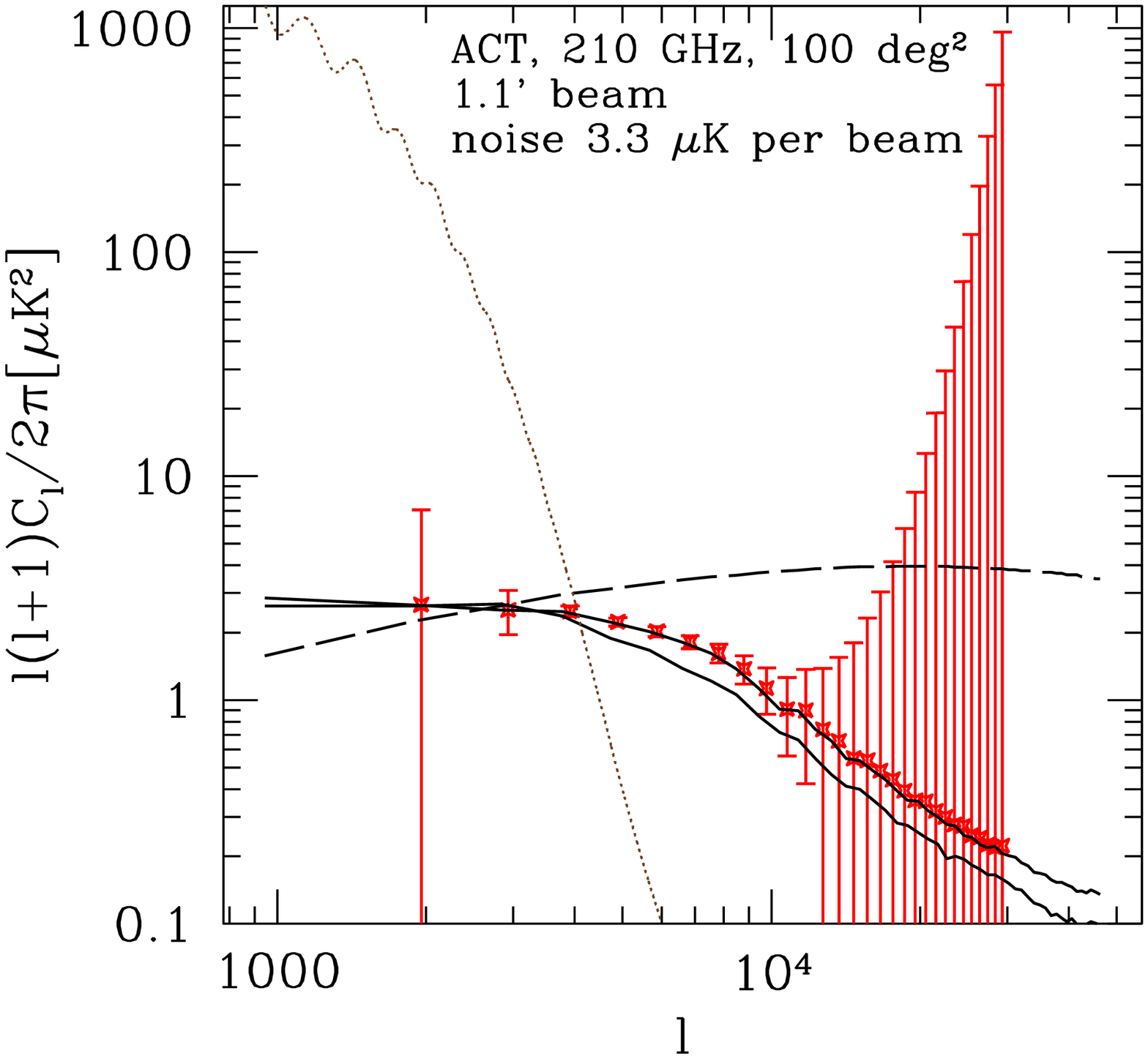}
  \includegraphics[width=3.3in]{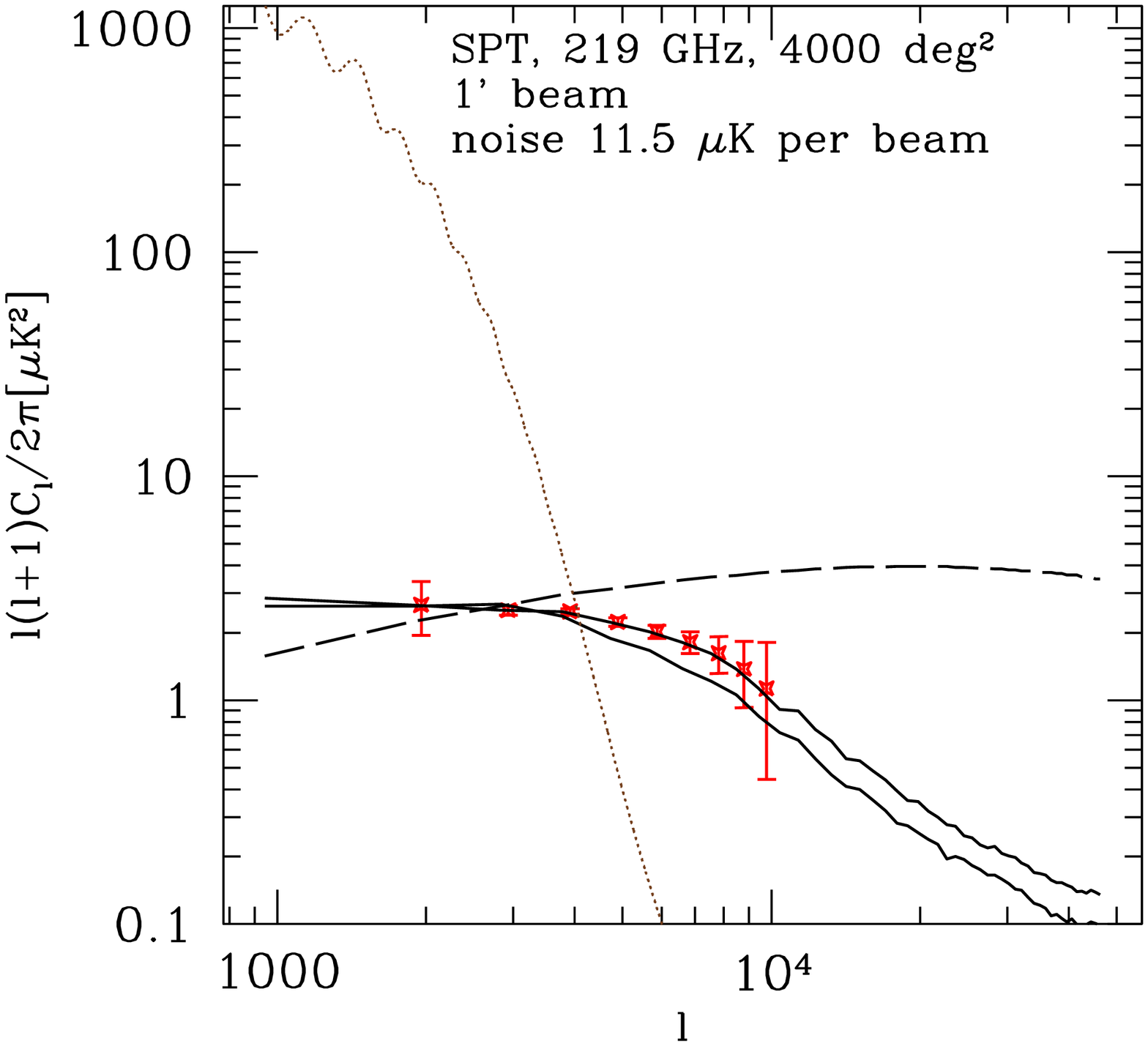}
\vspace{-0.6cm}
\caption{Observability of the kSZ: the sky power spectrum of the reionization 
signal (black, solid; simulations f250 and f250C) with the forecast error bars 
for ACT (left) and SPT (right) vs. $\ell$. The primary CMB anisotropy (dotted)
and the post-reionization kSZ signal (dashed) are also shown and are added to 
the noise error bars for the reionization signal. Cosmic variance from tSZ 
would increase the error bars, but here we are assuming that the tSZ component 
has been completely separated, by virtue of its characteristic spectral shape.
\label{act_spt}}
\end{center}
\end{figure*}
The Atacama Cosmology Telescope \citep{2004SPIE.5498....1F,2006astro.ph..8549K},
currently entering into operation, will observe at three frequency channels,
at 147, 215 and 279 GHz, targeting clear atmospheric windows, with bandwidths
23, 23 and 32~GHz and at resolutions 1.7', 1.3' and 0.9', respectively. The
target sensitivities in the three channels are 300, 500 and 700 $\mu\rm
K\,s^{-1/2}$, respectively, with a final aim of $\sim2\,\mu\,K$ per pixel over
a large area of the sky ($\sim200-400\,\rm deg^2$). The South Pole Telescope
(SPT) \citep{SPT} will be observing in 5 frequency channels, 95, 150, 219, 274
and 345~GHz, with similar bandwidths and resolution to ACT. Its sensitivity
might be even better, reaching $\sim10\,\mu$K over 1 deg$^2$ in an hour. Thus,
in terms of both resolution and sensitivity either bolometer is well-set to
detect the patchy reionization kSZ signal.  

The thermal noise of the detectors are given by 
\be
N_\ell=(sb)^2\exp\left[\frac{\ell(\ell+1)b^2}{8\ln2}\right]
\ee
assuming white noise with {\it rms} $s$ and a Gaussian beam with FWHM of
$b$. The error bar corresponding to a bin $\Delta\ell$ is then given by
\be
(\Delta C_{\ell})^2=\frac{2}{(2\ell+1)\Delta\ell f_{\rm
    sky}}(C_\ell+N_{\ell,tot})^2, 
\ee
where $C_\ell$ is our patchy reionization signal, $f_{\rm sky}$ is the sky 
coverage fraction of the survey and  
$N_{\ell,tot}=N_{\ell}+C_{\ell,primary}+C_{\ell,post-reion}$ is the 
average statistical noise for that bin. In the last expression we added
the primary and post-reionization signals to the noise, since for the purposes 
of patchy power spectrum measurement we assume that the primary CMB 
fluctuations are well normalized, and can be statistically subtracted, 
contributing to the statistical noise. Similarly, the post-reionization kSZ 
signal was forecast robustly by \citet{2004MNRAS.347.1224Z} and would be 
subtracted from the power spectra, but contributes to the statistical errors. 
In Figure~\ref{act_spt} we show our predicted kSZ signal for both of our WMAP3 
reionization models along with the ACT (left) and SPT (right) expected 
sensitivities, for $s=3.3\,\mu K$, $b=1.1'$, 100 deg$^2$ area (ACT; 
$f_{\rm sky}=100/41253=0.0024$; \citet{2005NewA...10..491H}) and 
$s=11.5\,\mu K$, $b=1'$, 4000 deg$^2$ area (SPT; $f_{\rm
  sky}=4000/41253=0.097$). This assumes perfect subtraction of all other
foregrounds. Results show that the reionization signal should be observable
with both ACT and SPT and in principle could even distinguish different
reionization scenarios. A number of difficult problems remain, however. 

The detected signal would be a mixture of thermal Sunyaev-Zel'dovich (tSZ) 
from galaxy clusters, gravitational lensing-induced anisotropies, Galactic 
dust and extragalactic point sources (e.g. dust in high redshift galaxies), 
in addition to the kSZ patchy reionization and post-reionization components. 
Separating these signals from each other presents significant challenges. The 
tSZ signal, which tends to dominate at these frequencies could be separated 
through its characteristic spectral shape, and in particular using the 
fact that the signal goes to zero at 217 GHz in the Earth frame. Detecting the
galaxy clusters with tSZ could allow also their kSZ contribution to be
evaluated and subtracted, through detailed modelling of the clusters based on
their tSZ data. The bright point sources could be subtracted based on
complementary observations with e.g. Atacama Large Millimetre Array
(ALMA). The lensing contribution is spectrally the same as the kSZ signal (and
the same as the primary CMB anisotropies), but is statistically-different from
the kSZ, which should allow their separation, at least in principle
\citep{Riquelme:2006nf}.  

The most difficult problem is to separate the patchy reionization and the 
post-reionization kSZ signals since both their spectra and their statistics 
are the same. Such separation is required, however if we want to extract 
any reionization information from the detected kSZ signal. This could be done 
for example by sufficiently detailed modelling of the post-reionization signal
and its properties. The linear effect, also called Ostriker-Vishniac (OV)
effect, can now be calculated with a reasonable precision, but it
significantly underestimates the actual post-reionization signal. The full
nonlinear kSZ  
post-reionization effect is still difficult to derive from simulations due to 
insufficient dynamic range. Current models and simulations roughly agree, but 
only within a factor of 2 at best, which would not allow for precise enough 
subtraction. Another option is to use the characteristic, fairly sharp peak 
of the patchy signal at $\ell$ of a few thousand, which is in contrast to the 
much broader peak of the post-reionization signal.

\section{Summary and Conclusions}
\label{conclusions_sect}
We presented detailed predictions of the signatures of inhomogeneous
reionization at redshifted 21-cm line of hydrogen and kSZ-induced CMB
small-scale anisotropies. Our results are based on the largest-scale radiative 
transfer simulations to date, utilizing a background cosmology given by the 
WMAP 3-year data. We discussed the observability of these signals in view of
the expected parameters and sensitivities of current and upcoming 21-cm and
kSZ experiments. We suggested some observational strategies based on our
results. In particular, the best approach for detecting the redshifted 21-cm
observations is to utilize relatively large beam sizes (a few arcminutes or
more) and bandwidths (hundreds of kHz), which would result in large gains in
flux, while retaining most of the signal. Additionally, it is better to
concentrate on the high frequencies, above 120 MHz, since the 21-cm
fluctuations, the corresponding fluxes and instrument sensitivities peak
there, while the foregrounds are noticeably lower than they are at lower
frequencies. 

%Discuss the dependence of these results on uncertainties (source
%efficiencies, clumping, etc.) and cosmology.

While these basic features of the reionization signals remain valid for any
reionization scenario in which most ionizing radiation is produced by stars in
galaxies, the detailed figures are dependent on the assumptions made about the
reionization parameters. These parameters are still highly uncertain and
include the ionizing source efficiencies (how many ionizing photons are
emitted per unit time) and gas clumpiness at small scales (how many
recombinations occur). It is difficult to estimate these parameters since the
simulations do not yet have the enormous dynamic range required and since we do
not understand sufficiently well the processes of star formation, galaxy
formation and escape of photons from galaxies. Instead, the approach we have
taken is to make simple, but reasonable assumptions for these parameters and
vary them within the range allowed by the current observational
constraints. As new and more detailed observations become available over time,
these will impose much more stringent constraints, which, combined with
further detailed simulations will be able to tell us more about the properties
of galaxies and stars at high redshifts. 

Another important caveat is that the results presented in this work are based
on  reionization simulations which do not resolve the smallest
atomically-cooling halos, with masses from $\sim10^8$ to
$\sim2\times10^9\,M_\odot$ and the even smaller molecularly-cooling
minihaloes. Smaller-box, higher-resolution radiative transfer simulations which
included all atomically-cooling halos \citep{2007MNRAS.376..534I} showed that
the presence of low-mass sources results in self-regulation of the
reionization process, whereby $\tau_{\rm es}$ is boosted, while the
large-scale structure of reionization and the 
epoch of overlap are largely unaffected. This is a consequence of the strong
suppression of these low-mass sources due to Jeans-mass filtering in the
ionized regions. We expect that this self-regulation would not affect our
current results significantly since the reionization signals discussed in this
work are dominated by the large bubbles. Utilizing smaller computational boxes 
in order to resolve the low-mass sources makes the problem more manageable, but 
results would underestimate the large-scale power of the ionization fluctuations
and be subject to a large cosmic variance. Resolving all halos of mass 
$\sim10^8M_\odot$ or larger in $100\,h^{-1}$~Mpc box would require $\sim10^{11}$ 
particles, with the additional complication that on such small scales
gasdynamical effects also become important and thus the complete treatment
would require a fully self-consistent N-body, gasdynamics and radiative
transfer. While still quite difficult, such simulations are now becoming
possible with the available algorithms and computer hardware. Future
higher-resolution calculations with more detailed microphysics would allow us
to evaluate more stringently the effects of low-mass sources and small-scale
structure on the reionization observables.  

\section*{Acknowledgments} 
We thank Wayne Hu and and Arthur Kosowsky for supplying the sensitivity data for 
SPT and ACT, respectively and for useful discussions. This work was partially 
supported by NASA Astrophysical Theory Program grants NAG5-10825 and NNG04G177G 
to PRS.


\begin{thebibliography}{53}
\expandafter\ifx\csname natexlab\endcsname\relax\def\natexlab#1{#1}\fi

\bibitem[{{Alvarez} {et~al.}(2006){Alvarez}, {Shapiro}, {Ahn}, \&
  {Iliev}}]{2006ApJ...644L.101A}
{Alvarez} M.~A., {Shapiro} P.~R., {Ahn} K., {Iliev} I.~T., 2006, \apjl, 644,
  L101

\bibitem[{{Barkana} \& {Loeb}(2005)}]{2005ApJ...626....1B}
{Barkana} R., {Loeb} A., 2005, \apj, 626, 1

\bibitem[{{Chuzhoy} {et~al.}(2006){Chuzhoy}, {Alvarez}, \&
  {Shapiro}}]{2006ApJ...648L...1C}
{Chuzhoy} L., {Alvarez} M.~A., {Shapiro} P.~R., 2006, \apjl, 648, L1

\bibitem[{{Dore} {et~al.}(2007){Dore}, {Holder}, {Alvarez}, {Iliev}, {Mellema},
  {Pen}, \& {Shapiro}}]{2007astro.ph..1784D}
{Dore} O., {Holder} G., {Alvarez} M., {Iliev} I.~T., {Mellema} G., {Pen} U.-L.,
  {Shapiro} P.~R., 2007, Phys.Rev.D submitted, (astro-ph/0701784)

\bibitem[{{Field}(1959)}]{1959ApJ...129..536F}
{Field} G.~B., 1959, \apj, 129, 536

\bibitem[{{Fowler}(2004)}]{2004SPIE.5498....1F}
{Fowler} J.~W., 2004, in Z-Spec: a broadband millimeter-wave grating
  spectrometer: design, construction, and first cryogenic measurements.
  Proceedings of the SPIE, Volume 5498, pp. 1-10 (2004)., {Bradford} C.~M.,
  {Ade} P.~A.~R., {Aguirre} J.~E., {Bock} J.~J., {Dragovan} M., {Duband} L.,
  {Earle} L., {Glenn} J., {Matsuhara} H., {Naylor} B.~J., {Nguyen} H.~T., {Yun}
  M., {Zmuidzinas} J., eds., pp. 1--10

\bibitem[{{Furlanetto}(2006)}]{2006MNRAS.371..867F}
{Furlanetto} S.~R., 2006, \mnras, 371, 867

\bibitem[{{Furlanetto} \& {Loeb}(2002)}]{2002ApJ...579....1F}
{Furlanetto} S.~R., {Loeb} A., 2002, \apj, 579, 1

\bibitem[{{Furlanetto} \& {Loeb}(2004)}]{2004ApJ...611..642F}
---, 2004, \apj, 611, 642

\bibitem[{{Furlanetto} {et~al.}(2006){Furlanetto}, {Oh}, \&
  {Briggs}}]{2006PhR...433..181F}
{Furlanetto} S.~R., {Oh} S.~P., {Briggs} F.~H., 2006, \physrep, 433, 181

\bibitem[{{Gnedin} \& {Jaffe}(2001)}]{2001ApJ...551....3G}
{Gnedin} N.~Y., {Jaffe} A.~H., 2001, \apj, 551, 3

\bibitem[{{Gruzinov} \& {Hu}(1998)}]{1998ApJ...508..435G}
{Gruzinov} A., {Hu} W., 1998, \apj, 508, 435

\bibitem[{{Holder} {et~al.}(2006){Holder}, {Iliev}, \& {Mellema}}]{pol21}
{Holder} G., {Iliev} I.~T., {Mellema} G., 2006, ApJ, submitted,
  (astro-ph/0609689)

\bibitem[{{Hu}(2000)}]{2000ApJ...529...12H}
{Hu} W., 2000, \apj, 529, 12

\bibitem[{{Huffenberger} \& {Seljak}(2005)}]{2005NewA...10..491H}
{Huffenberger} K.~M., {Seljak} U., 2005, New Astronomy, 10, 491

\bibitem[{{Iliev} {et~al.}(2006{\natexlab{a}}){Iliev}, {Ciardi}, {Alvarez},
  {Maselli}, {Ferrara}, {Gnedin}, {Mellema}, {Nakamoto}, {Norman}, {Razoumov},
  {Rijkhorst}, {Ritzerveld}, {Shapiro}, {Susa}, {Umemura}, \&
  {Whalen}}]{comparison1}
{Iliev} I.~T., {Ciardi} B., {Alvarez} M.~A., {Maselli} A., {Ferrara} A.,
  {Gnedin} N.~Y., {Mellema} G., {Nakamoto} T., {Norman} M.~L., {Razoumov}
  A.~O., {Rijkhorst} E.-J., {Ritzerveld} J., {Shapiro} P.~R., {Susa} H.,
  {Umemura} M., {Whalen} D.~J., 2006{\natexlab{a}}, \mnras, 371, 1057

\bibitem[{{Iliev} {et~al.}(2006{\natexlab{b}}){Iliev}, {Mellema}, {Pen},
  {Merz}, {Shapiro}, \& {Alvarez}}]{2006MNRAS.369.1625I}
{Iliev} I.~T., {Mellema} G., {Pen} U.-L., {Merz} H., {Shapiro} P.~R., {Alvarez}
  M.~A., 2006{\natexlab{b}}, \mnras, 369, 1625

\bibitem[{{Iliev} {et~al.}(2007{\natexlab{a}}){Iliev}, {Mellema}, {Shapiro}, \&
  {Pen}}]{2007MNRAS.376..534I}
{Iliev} I.~T., {Mellema} G., {Shapiro} P.~R., {Pen} U.-L., 2007{\natexlab{a}},
  \mnras, 376, 534

\bibitem[{{Iliev} {et~al.}(2007{\natexlab{b}}){Iliev}, {Pen}, {Bond},
  {Mellema}, \& {Shapiro}}]{kSZ}
{Iliev} I.~T., {Pen} U.-L., {Bond} J.~R., {Mellema} G., {Shapiro} P.~R.,
  2007{\natexlab{b}}, ApJ, submitted, ArXiv Astrophysics e-prints
  (astro-ph/0609592), 660, 933

\bibitem[{{Iliev} {et~al.}(2006{\natexlab{c}}){Iliev}, {Pen}, {Richard Bond},
  {Mellema}, \& {Shapiro}}]{2006NewAR..50..909I}
{Iliev} I.~T., {Pen} U.-L., {Richard Bond} J., {Mellema} G., {Shapiro} P.~R.,
  2006{\natexlab{c}}, New Astronomy Review, 50, 909

\bibitem[{{Iliev} {et~al.}(2003){Iliev}, {Scannapieco}, {Martel}, \&
  {Shapiro}}]{2003MNRAS.341...81I}
{Iliev} I.~T., {Scannapieco} E., {Martel} H., {Shapiro} P.~R., 2003, \mnras,
  341, 81

\bibitem[{{Iliev} {et~al.}(2002){Iliev}, {Shapiro}, {Ferrara}, \&
  {Martel}}]{2002ApJ...572L.123I}
{Iliev} I.~T., {Shapiro} P.~R., {Ferrara} A., {Martel} H., 2002, \apjl, 572,
  L123

\bibitem[{{Jaffe} \& {Kamionkowski}(1998)}]{1998PhRvD..58d3001J}
{Jaffe} A.~H., {Kamionkowski} M., 1998, \prd, 58, 043001

\bibitem[{{Kosowsky} \& {the ACT Collaboration}(2006)}]{2006astro.ph..8549K}
{Kosowsky} A., {the ACT Collaboration}, 2006, ArXiv Astrophysics e-prints
  (astro-ph/0608549)

\bibitem[{{Kuo} \& {et al.}(2006)}]{acbar}
{Kuo} C.~L., {et al.}, 2006, ApJ, submitted (astro-ph/0611198)

\bibitem[{{Ma} \& {Fry}(2002)}]{2002PhRvL..88u1301M}
{Ma} C.-P., {Fry} J.~N., 2002, Physical Review Letters, 88, 211301

\bibitem[{{Madau} {et~al.}(1997){Madau}, {Meiksin}, \&
  {Rees}}]{1997ApJ...475..429M}
{Madau} P., {Meiksin} A., {Rees} M.~J., 1997, \apj, 475, 429

\bibitem[{{McQuinn} {et~al.}(2005){McQuinn}, {Furlanetto}, {Hernquist}, {Zahn},
  \& {Zaldarriaga}}]{2005ApJ...630..643M}
{McQuinn} M., {Furlanetto} S.~R., {Hernquist} L., {Zahn} O., {Zaldarriaga} M.,
  2005, \apj, 630, 643

\bibitem[{{Mellema} {et~al.}(2006{\natexlab{a}}){Mellema}, {Iliev}, {Alvarez},
  \& {Shapiro}}]{methodpaper}
{Mellema} G., {Iliev} I.~T., {Alvarez} M.~A., {Shapiro} P.~R.,
  2006{\natexlab{a}}, New Astronomy, 11, 374

\bibitem[{{Mellema} {et~al.}(2006{\natexlab{b}}){Mellema}, {Iliev}, {Pen}, \&
  {Shapiro}}]{2006MNRAS.372..679M}
{Mellema} G., {Iliev} I.~T., {Pen} U.-L., {Shapiro} P.~R., 2006{\natexlab{b}},
  \mnras, 372, 679

\bibitem[{{Merz} {et~al.}(2005){Merz}, {Pen}, \& {Trac}}]{2005NewA...10..393M}
{Merz} H., {Pen} U.-L., {Trac} H., 2005, New Astronomy, 10, 393

\bibitem[{{Morales} \& {Hewitt}(2004)}]{2004ApJ...615....7M}
{Morales} M.~F., {Hewitt} J., 2004, \apj, 615, 7

\bibitem[{{Ostriker} \& {Vishniac}(1986)}]{1986ApJ...306L..51O}
{Ostriker} J.~P., {Vishniac} E.~T., 1986, \apjl, 306, L51

\bibitem[{{Pritchard} \& {Furlanetto}(2006)}]{2006astro.ph..7234P}
{Pritchard} J.~R., {Furlanetto} S.~R., 2006, ArXiv Astrophysics e-prints
  (astro-ph/0607234)

\bibitem[{Riquelme \& Spergel(2006)}]{Riquelme:2006nf}
Riquelme M.~A., Spergel D.~N., 2006, ArXiv Astrophysics e-prints
  (astro-ph/0610007)

\bibitem[{Ruhl \& {et al.}(2006)}]{SPT}
Ruhl J.~E., {et al.}, 2006, Proc. SPIE, 5498, 11

\bibitem[{{Salvaterra} {et~al.}(2005){Salvaterra}, {Ciardi}, {Ferrara}, \&
  {Baccigalupi}}]{2005MNRAS.360.1063S}
{Salvaterra} R., {Ciardi} B., {Ferrara} A., {Baccigalupi} C., 2005, \mnras,
  360, 1063

\bibitem[{{Santos} {et~al.}(2003){Santos}, {Cooray}, {Haiman}, {Knox}, \&
  {Ma}}]{2003ApJ...598..756S}
{Santos} M.~G., {Cooray} A., {Haiman} Z., {Knox} L., {Ma} C.-P., 2003, \apj,
  598, 756

\bibitem[{{Scott} \& {Rees}(1990)}]{1990MNRAS.247..510S}
{Scott} D., {Rees} M.~J., 1990, \mnras, 247, 510

\bibitem[{{Seljak} {et~al.}(2006){Seljak}, {Slosar}, \&
  {McDonald}}]{2006astro.ph..4335S}
{Seljak} U., {Slosar} A., {McDonald} P., 2006, ArXiv Astrophysics e-prints
  (astro-ph/0604335)

\bibitem[{{Sethi}(2005)}]{2005MNRAS.363..818S}
{Sethi} S.~K., 2005, \mnras, 363, 818

\bibitem[{{Shapiro} {et~al.}(2006){Shapiro}, {Ahn}, {Alvarez}, {Iliev},
  {Martel}, \& {Ryu}}]{2006ApJ...646..681S}
{Shapiro} P.~R., {Ahn} K., {Alvarez} M.~A., {Iliev} I.~T., {Martel} H., {Ryu}
  D., 2006, \apj, 646, 681

\bibitem[{{Shaver} {et~al.}(1999){Shaver}, {Windhorst}, {Madau}, \& {de
  Bruyn}}]{1999A&A...345..380S}
{Shaver} P.~A., {Windhorst} R.~A., {Madau} P., {de Bruyn} A.~G., 1999, \aap,
  345, 380

\bibitem[{{Spergel} \& {et al.}(2003)}]{2003ApJS..148..175S}
{Spergel} D.~N., {et al.}, 2003, \apjs, 148, 175

\bibitem[{{Spergel} \& {et al.}(2006)}]{2006astro.ph..3449S}
---, 2006, ArXiv Astrophysics e-prints (arXiv:astro-ph/0603449)

\bibitem[{{Springel} {et~al.}(2001){Springel}, {White}, \&
  {Hernquist}}]{2001ApJ...549..681S}
{Springel} V., {White} M., {Hernquist} L., 2001, \apj, 549, 681

\bibitem[{{Sunyaev} \& {Zeldovich}(1980)}]{1980MNRAS.190..413S}
{Sunyaev} R.~A., {Zeldovich} I.~B., 1980, \mnras, 190, 413

\bibitem[{{Vishniac}(1987)}]{1987ApJ...322..597V}
{Vishniac} E.~T., 1987, \apj, 322, 597

\bibitem[{{Yao} \& {et al.}(2006)}]{2006JPhG...33....1Y}
{Yao} W.-M., {et al.}, 2006, Journal of Physics G Nuclear Physics, 33, 1

\bibitem[{{Zahn} {et~al.}(2005){Zahn}, {Zaldarriaga}, {Hernquist}, \&
  {McQuinn}}]{2005ApJ...630..657Z}
{Zahn} O., {Zaldarriaga} M., {Hernquist} L., {McQuinn} M., 2005, \apj, 630, 657

\bibitem[{{Zaldarriaga} {et~al.}(2004){Zaldarriaga}, {Furlanetto}, \&
  {Hernquist}}]{2004ApJ...608..622Z}
{Zaldarriaga} M., {Furlanetto} S.~R., {Hernquist} L., 2004, \apj, 608, 622

\bibitem[{{Zeldovich} \& {Sunyaev}(1969)}]{1969Ap&SS...4..301Z}
{Zeldovich} Y.~B., {Sunyaev} R.~A., 1969, \apss, 4, 301

\bibitem[{{Zhang} {et~al.}(2004){Zhang}, {Pen}, \&
  {Trac}}]{2004MNRAS.347.1224Z}
{Zhang} P., {Pen} U.-L., {Trac} H., 2004, \mnras, 347, 1224

\end{thebibliography}
\end{document}